\documentclass{ws-ijqi}
\renewcommand{\trimmarks}{\relax}

\newcommand{\JournalReference}{%
\raisebox{6ex}{\makebox[0pt][l]{\normalfont\small\hspace{2.5em}%
International Journal of Quantum Information \textbf{1} (2003) 153--188}}}

\renewcommand{\draftnote}{\centerline%
{\underline{\texttt{"\jobname.tex", draft of  \today, \currenttime}}}}

\renewcommand{\draftnote}{\relax}

\def\currenttime{\hour=\time \divide\hour by 60 \number\hour:%
  \multiply\hour by 60 \minute=\time \global\advance\minute by -\hour%
  \ifnum\minute<10 0\number\minute\else\number\minute\fi}

\newcommand{\I}{\mathrm{i}}
\newcommand{\ket}[1]{\bigl|#1\bigr\rangle}
\newcommand{\bra}[1]{\bigl\langle #1\bigr|}
\newcommand{\ketbra}[2]{\bigl|#1\bigr\rangle\bigl\langle #2\bigr|}
\newcommand{\proj}[1]{\ketbra{#1}{#1}}
\newcommand{\braket}[2]{\bigl\langle #1\bigr|#2\bigr\rangle}
\newcommand{\expect}[1]{\bigl\langle #1\bigr\rangle}
\newcommand{\Exp}[1]{{\rm e}^{\mbox{\footnotesize$#1$}}}
\newcommand{\power}[1]{^{\mbox{\footnotesize$#1$}}}
\newcommand{\barr}[1]{\overline{#1}}
\newcommand{\adj}{^{\dagger}}
\newcommand{\phadj}{^{\phantom{\dagger}}}
\newcommand{\phstar}{^{\phantom{*}}}
\newcommand{\repr}{\widehat{=}}
\newcommand{\matr}[1]{\left[\begin{array}{cc} #1 \end{array}\right]}
\newcommand{\spmatr}[2][\quad]%
{\left[\begin{array}{c@{#1}c} #2 \end{array}\right]}
\newcommand{\Matr}[1]{\left[\begin{array}{cccc} #1 \end{array}\right]}
\newcommand{\col}[1]{\left[\begin{array}{l} #1 \end{array}\right]}
\newcommand{\row}[1]{\left[ #1 \right]}
\newcommand{\Tr}[2][]{\mathrm{Tr}_{#1}\left\{ #2 \right\}}

\newcommand{\half}{\frac{1}{2}}
\newcommand{\thalf}{\tfrac{1}{2}}
\newcommand{\trans}{^{\mathrm{T}}}
\newcommand{\ptrans}{^{\mathrm{T}_1}}
\newcommand{\mq}{\mathsf{q}}
\newcommand{\mx}{\mathsf{x}}
\newcommand{\mz}{\mathsf{z}}

\newcommand{\ma}{\mathsf{a}}
\newcommand{\mA}{\mathsf{A}}
\newcommand{\mI}{\mathsf{I}}

\newcommand{\mW}{\mathsf{W}}
\newcommand{\mE}{\mathsf{E}}

\newcommand{\mQ}{\mathsf{Q}}
\newcommand{\mC}{\mathsf{C}}
\newcommand{\mU}{\mathsf{U}}
\newcommand{\mT}{\mathsf{T}}
\newcommand{\mP}{\mathsf{P}}
\newcommand{\mD}{\mathsf{D}}
\newcommand{\Real}{\mathop{\mathrm{Re}}}
\newcommand{\Imag}{\mathop{\mathrm{Im}}}
\newcommand{\D}{\mathrm{d}}
\newcommand{\tG}{\widetilde{G}}

\newcommand{\tmQ}{\widetilde{\mQ}}

\newcommand{\JournalTitle}[1]{\textit{#1}\ }
\newcommand{\PR}{\JournalTitle{Phys.\ Rev.}}
\newcommand{\PRA}{\JournalTitle{Phys.\ Rev.\ A}}
\newcommand{\PRL}{\JournalTitle{Phys.\ Rev.\ Lett.}}

\newcommand{\ZPhys}{\JournalTitle{Z.~Phys.}}
\newcommand{\PNAS}{\JournalTitle{Proc.\ Natl.\ Acad.\ Sci.\ U.~S.~A.}}
\newcommand{\PCamPS}{\JournalTitle{Proc.\ Camb.\ Phil.\ Soc.}}
\newcommand{\PhRep}{\JournalTitle{Phys.\ Rep.}}
\newcommand{\JPhysA}{\JournalTitle{J. Phys.\ A: Math.\ Gen.}}
\newcommand{\OC}{\JournalTitle{Opt.\ Commun.}}
\newcommand{\JMO}{\JournalTitle{J. Mod.\ Opt.}}

\usepackage{graphicx} 

\begin{document}

\markboth{B.-G.~Englert and K.~W\'odkiewicz}
{Tutorial Notes on One-Party and Two-Party Gaussian States}


\title{\JournalReference%
\uppercase{Tutorial Notes on One-Party and Two-Party Gaussian States}}

\author{\uppercase{Berthold-Georg Englert}}
\address{Department of Physics, %
National University of Singapore, Singapore 117542\\
phyebg@nus.edu.sg}

\author{\uppercase{Krzysztof W\'odkiewicz}}
\address{Instytut Fizyki Teoretycznej, Uniwersytet Warszawski,
 Warszawa 00--681, Poland\\
and\\
Department of Physics and Astronomy, University of New Mexico\\
 Albuquerque, NM 87131, USA\\
Krzysztof.Wodkiewicz@fuw.edu.pl}

\maketitle

\begin{history}
\received{28 July 2003}
\end{history}

\begin{abstract}
Gaussian states --- or, more generally, Gaussian operators --- 
play an important role in Quantum Optics and Quantum Information Science, 
both in discussions about conceptual issues and in practical applications.
We describe, in a tutorial manner, a systematic operator method for first
characterizing such states and then investigating their properties.
The central numerical quantities are the covariance matrix that specifies the
characteristic function of the state, and the closely related matrices
associated with Wigner's and Glauber's phase space functions.
For pedagogical reasons, we restrict the discussion to one-dimensional and 
two-dimensional Gaussian states, for which we provide illustrating and
instructive examples.
\end{abstract}



\section{Introduction}

Gaussian wave functions appeared already very early in the development of
Quantum Mechanics, when Schr\"odigner explained the observed  behavior 
of the linear harmonic oscillator in terms of the undulatory eigenfunctions 
of  the corresponding differential equation.\cite{Schroed27} 
Heisenberg, in his fundamental work of 1927 devoted to the
uncertainty relation,\cite{Heis27} used a Gaussian wave function of the form
(in historical notation) 
\begin{equation}
 \label{gauss}S(\eta,q)\, \  \mathrm{prop}\
 \, \Exp{-\frac{(q-q^{\prime})^2}{2q_{1}^2}
-\frac{2\pi\I}{h}p^{\prime}(q-q^{\prime})}
\end{equation}
to show that the undulatory character of this wave function,%
\footnote{In Heisenberg's more general context, this Gaussian wave function
refers to an arbitrary one-dimensional system, be it a free particle, a
harmonic oscillator, or something with other dynamical properties. 
For the sake of convenience, we shall invariably speak of harmonic
oscillators, thinking in particular of those associated with modes of the
quantized radiation field.}\ 
combined with the Dirac-Jordan transformation theory, leads to the
indeterminacy relations. 
Heisenberg used this Gaussian  wave function as a probability amplitude
--- perhaps the first application of this kind --- to exhibit, in his own
words, that
\emph{The more precisely the position is determined, the less
precisely the momentum is known in this instant, and vice versa}.

Gaussian statistical properties are fundamental to many applications in
statistical physics. 
It is familiar that the classical theory of  a Gaussian noise associated with
Brownian motion  is fully characterized by the covariance function of the phase
space variables $q$ and $p$. 
The transition from classical phase space variables to
canonical quantum  position and momentum operators requires ``quantization
rules'' for associating quantum operator functions $F(q,p)$ with the
classical numerical functions $\mathcal{F}(q,p)$. 

In the framework of quantum noise associated with  a one-dimensional harmonic
oscillator, Gardiner\cite{Ga91} was perhaps the first to discuss, in a textbook
format, various phase space properties of the most general Gaussian
operators. 
Such a systematic formalism is very useful 
for the discussion of Gaussian statistical
properties of a single quantum system with one degree of freedom.

The remarkable role that is played by one-dimensional Gaussian wave functions
in realizing the ultimate limit of Heisenberg's uncertainty relation for a
single position-momentum pair naturally invites generalizations to 
higher-dimensional systems.
That, however, opened a path into the then-unexplored territory of quantum
correlations and quantum entanglement and thus started 
a never ending story.

In their reasoning concerning the alleged incompleteness of
quantum mechanics,\cite{EPR35} Einstein, Podolsky, and Rosen (EPR)
used in 1935 the following wave function for a system composed of
two particles (in historical notation):
\begin{equation}
\label{epr} \Psi (x_{1},x_{2})= \int_{-\infty}^{\infty}\D p\
\Exp{(2\pi\I/h)(x_1-x_2+x_0)p}\;.
\end{equation}
The fact that this function could be written as a sum of products of one
factor each for the two particles,
\begin{equation}
\label{entan}
 \Psi (x_{1},x_{2})=
\int_{-\infty}^{\infty}\D p\ \psi_p(x_1) u_p(x_2)\,,
\end{equation}
leads to the intriguing concept of quantum \emph{entanglement}, a term coined
by Schr\"odinger in Refs.~\refcite{Schroed35a} and \refcite{Schroed35b}.%
\footnote{It was in Ref.~\refcite{Schroed35a} 
that the word \emph{entanglement}, which refers to the non-separability 
of quantum states of composite systems, was given its
familiar meaning within the context of quantum mechanics. 
For a brief history about the use of the English word \emph{entanglement} 
and the German word \emph{Verschr\"ankung} see Ref.~\refcite{ekert}.}

The entangled EPR wave function (\ref{epr}) is a singular function
of the distance $x_1-x_2$ and can be visualized as an infinitely
sharp two-dimensional Gaussian wave function of the entangled 
two-party system. 
Soon after the publication of the EPR paper, Bohr
pointed out\cite{bohr35} that the two pairs of canonical
variables $q_1,p_1$ and $q_2,p_2$ for the two particles of a composed
system can be replaced by two new pairs of conjugate variables,
\begin{equation}
\label{bohr} 
\mathcal{Q}_{1,2}= \frac{1}{\sqrt{2}}(q_1 \mp q_2)
\,,\quad 
\mathcal{P}_{1,2}= \frac{1}{\sqrt{2}}(p_1 \mp p_2) \,,
\end{equation}
each pair now referring to both particles.
Bohr noted that, since they commute: 
$\mathcal{Q}_1\mathcal{P}_2=\mathcal{P}_2\mathcal{Q}_1$, 
the observables $\mathcal{Q}_1$ and $\mathcal{P}_2$ can be assigned 
sharp values simultaneously, so that the wave
function (\ref{epr}) represents a joint eigenstate  of these
commuting variables. 
This property is at the heart of quantum entanglement, 
unintentionally brought to light by EPR.
Bell inequalities of some kind are violated for the EPR wave function, 
as can be
demonstrated by using its Wigner representation.\cite{BaWo98}

Schr\"odinger, Heisenberg, and the EPR trio were dealing with systems 
described by wave functions, i.e., with pure quantum states. 
For a quantum system described by a Gaussian mixed statistical operator
$\rho$, the correspondence between classical Gaussian functions
$\mathcal{G}(q_1,p_1,q_2,p_2)$ 
in two dimensions and the statistical operators $\rho=G(q_1,p_1,q_2,p_2)$
should employ a ``quantization rule'' that preserves 
all the properties of  a density matrix.

Recent applications of entangled two-mode squeezed states of light for
quantum teleportation\cite{Furusawa+al98} and other quantum
information purposes\cite{Ou+al92} have generated a lot of
interest in the entangled properties of general mixed Gaussian
states in quantum optics.\cite{Braunstein+al01}
It turns out that the concept of quantum entanglement, as defined
by (\ref{entan}), has to be generalized when the system 
is not in a pure state.
In the general case of a density operator, rather than a wave function, 
one uses the definition of quantum \emph{separability}
introduced by Werner:\cite{Werner89} 
A general quantum density operator of a two-party system is separable if it
is a convex sum of product states,
\begin{equation}
\label{separability}
 \rho = \sum_k w_k\,\rho_1^{(k)}\,\rho_2^{(k)}\quad
\mathrm{with} \quad \sum_k w_k=1 \quad \mathrm{and} \quad w_k>0
\,,
\end{equation}
where $\rho_1^{(k)}$ and $\rho_2^{(k)}$ are statistical operators
of the two subsystems in question.%
\footnote{Actually, one should only require that the given $\rho$ 
can be approximated to any required accuracy by a sum of this kind 
with a finite number of terms, but we take the liberty to ignore 
mathematical details of such a more pedantic sort.} 

The separability properties of general mixed states of a harmonic oscillator in
two dimensions can be studied with the help of two different methods. 
The first uses the covariance matrix and the Heisenberg uncertainty 
relations.\cite{Duan+al00} 
The second makes use of the criterion of positivity under
partial transposition.\cite{Simon00} 
These two approaches lead to essentially
the same conclusions, while employing very different techniques.

The objective of this tutorial is to review different operator and phase-space
techniques that allows a systematic investigation of different properties of
Gaussian operators of unit trace, referring, for instance, to a single mode or
two modes of the radiation field. 
We discuss the properties of Gaussian operators with the aid of 
various techniques that are widely used in Quantum Optics.\cite{QO:tools}
In particular, we provide a careful description of Gaussian operators using
(i)~the boson operator algebra, and (ii)~phase-space descriptions 
based on the Wigner representation and the Glauber P-representation. 
We establish relations between these equivalent though different-in-form 
versions of Gaussian statistical operators. 
In  Sec.~\ref{sec:1D} we gather the necessary mathematical tools used
in this tutorial at the example of the general Gaussian operator of a
one-dimensional harmonic oscillator. 
We provide a general parametrization of such states, 
discuss the positivity criteria and the so-called P-representability of
positive Gaussian operators. 
At the end of Sec.~\ref{sec:1D} we illustrate the 
formalism with the example of a squeezed single-mode state of light. 
Then, in Sec.~\ref{sec:2D}, we turn to two-party Gaussian states.
The central part of this tutorial is Sec.~\ref{sec:2D-SepCrit} 
that deals with the separability of two-dimensional Gaussian states. 
We discuss the positivity criteria of such states and their separability 
conditions, thereby illustrating
various approaches. 
At the end of Sec.~\ref{sec:2D} we provide a
number of simple examples that are useful for studying and illustrating 
various physical effects in Quantum Optics\cite{QO:gen} 
and in Quantum Information Theory.\cite{QI:gen} 
We confine our discussion to two-party systems, but the methods presented 
in this tutorial can be extended to many-party systems.
In concluding remarks we summarize and suggest some supplementary reading on
the subject.


\section{Gaussian States of a One-Dimensional Oscillator}\label{sec:1D}

\subsection{Parameterizations}
Any operator referring to a harmonic oscillator --- position operator $q$,
momentum operator $p$, both measured in natural units, so that $qp-pq=\I$ ---
is a function of the familiar ladder operators
\begin{equation}
  a\adj=\frac{q-\I p}{\sqrt{2}}\,,\quad a=\frac{q+\I p}{\sqrt{2}}\,.
\end{equation}
We can specify such an operator $G(a\adj,a)$ by its
characteristic function $C(z^*,z)$,
\begin{equation}
  C(z^*,z)=\Tr{\Exp{z a\adj-z^*a}G(a\adj,a)}\,,
\label{eq:GtoC}
\end{equation}
which is a numerical function of the complex phase space variables
\begin{equation}
  \label{eq:PhSpVariables}
    z^*=\frac{q'-\I p'}{\sqrt{2}}\,,\quad z=\frac{q'+\I p'}{\sqrt{2}}\,.
\end{equation}
Here, $q',p'$ are the cartesian coordinates of classical phase space as one
knows them from Hamilton's approach to classical mechanics or the Liouville
formulation of statistical mechanics.
A more compact way of writing (\ref{eq:GtoC}) is
\begin{equation}
  C(\mz)=\Tr{\Exp{-\mz\adj\mE\ma}G(\ma)}
\label{eq:GtoC'}
\end{equation}
where we introduce 2-component columns and rows in accordance with
\begin{equation}
  \mz=\col{z\\z^*}\,,\ \mz\adj=\row{z^*,z}\,,\quad
  \ma=\col{a\\a\adj}\,,\ \ma\adj=\row{a\adj,a}\,,
  \label{eq:zadef}
\end{equation}
and meet the symplectic $2\times2$ matrix
\begin{equation}
\label{eq:defE}
\mE=\matr{1&0\\0&-1}\,.
\end{equation}
Reciprocally, we get the operator from its characteristic function by a
phase space integration,
\begin{equation}
  \label{eq:CtoG}
  G(\ma)=\int\!(\D\mz)\,\Exp{\mz\adj\mE\ma}C(\mz)\,,
\end{equation}
which is essentially the Liouville trace of statistical mechanics and
complements the quantum-mechanical trace of (\ref{eq:GtoC'}).
The explicit form of the volume element $(\D\mz)$ depends on the way
phase space is parameterized.
For the standard parameterization (\ref{eq:PhSpVariables}) one has
\begin{equation}
  (\D\mz)=\frac{\D q'\,\D p'}{2\pi}
\end{equation}
or $(\D\mz)=\frac{1}{\pi}\D\Real z\;\D\Imag z=\frac{1}{\pi}\D^2z$
in a popular notation.
But sometimes the actual integrations are more easily performed with
\begin{equation}
  \label{eq:dz2}
  z=q'\,,\quad z^*=-\I p'\,,\quad
   (\D\mz)=\frac{\D q'\,\D p'}{2\pi}\,,
\end{equation}
for instance.
Note that we include the normalizing denominator
of $2\pi$ in $(\D\mz)$ which, in a rough
manner of speaking, indicates that the integral in (\ref{eq:CtoG})
counts one quantum state per phase space volume of $2\pi\repr2\pi\hbar$.

We focus on Hermitian operators $G(\ma)=\bigl[G(\ma)\bigr]\adj$
with characteristic functions that are of the Gaussian form
\begin{equation}
   C(\mz) =  \Exp{-\half\mz\adj\mC\mz}
=\Exp{-\half m{z^*}^2-(n+\half)z^*z-\half m^*z^2}
\label{eq:Cdef}
\end{equation}
with the $2\times2$ matrix $\mC$ given by
\begin{equation}
 \mC=\matr{n+\half & m \\ m^* &n+\half}=\mC\adj\,.
  \label{eq:mCdef}
\end{equation}
The hermiticity of $G$ implies
$\bigl[C(\mz)\bigr]^*=C(-\mz)$, and vice versa;
the Gaussian characteristic function (\ref{eq:Cdef}) has this symmetry
property because the parameter $n$ is real.

For a given $C(\mz)$, (\ref{eq:Cdef}) does not specify the diagonal entries of
$\mC$ uniquely, only their sum is determined.
The symmetry of the standard form (\ref{eq:mCdef}) of $\mC$ exploits
this arbitrariness conveniently.
A particularly useful way of looking at this proceeds from the response of
$\ma$ and $\mz$ to transposition,
\begin{equation}
  \ma\trans=\row{a,a\adj}=\ma\adj\mT\,,\quad
  \mz\trans=\row{z,z^*}=\mz\adj\mT\,,
\end{equation}
with
\begin{equation}
  \label{eq:mTdef}
  \mT=\matr{0&1\\1&0}\,,
\end{equation}
so that
\begin{equation}
  \left(\mz\adj\mC\mz\right)\trans=\mz\adj\mT\mC\trans\mT\mz\,.
\end{equation}
This states that one can replace $\mC$ by $\mT\mC\trans\mT$ in
(\ref{eq:Cdef}),
or by any linear average of both, without changing the characteristic function
$C(\mz)$.
It is therefore natural to enforce the symmetry
\begin{equation}
  \label{eq:Csymmetry}
  \mC=\mT\mC\trans\mT\,,
\end{equation}
and this establishes (\ref{eq:mCdef}).

Since $C(\mz)=1$ for $z^*=0$, $z=0$ in (\ref{eq:Cdef})
we take for granted that $G$ has unit trace, $\Tr{G}=1$.
This excludes Gaussian operators with infinite trace, but otherwise it is
just a matter of convenient conventional normalization.

The absence of linear terms in the exponent indicates another convention:
We assume that $\Tr{aG}=0$, $\Tr{a\adj G}=0$, which can always be arranged
with a suitable unitary shift of $a\adj$ and $a$.

Upon expanding the characteristic function in powers of $z^*$ and
$z$ one easily identifies the physical significance of the numerical
parameters $n$ and $m$,
\begin{eqnarray}
  &n=\Tr{a\adj aG}\,,&
\nonumber\\
&m=-\Tr{a^2G}\,,\quad m^*=-\Tr{{a\adj}^2G}\,,&
\end{eqnarray}
or, more compactly,
\begin{equation}
  \label{eq:n+m'}
 \Tr{\ma\ma\adj G}=\mE\mC\mE+\thalf\mE
\end{equation}
and
\begin{equation}
  \mC=\mE\,\Tr{\ma\ma\adj G}\mE-\thalf\mE
=\thalf\mE\,
\Tr{\bigl[\ma\ma\adj+\mT\bigl(\ma\ma\adj\bigr)\trans\mT\bigr]G}\mE
\;.
\end{equation}
These relations identify $\mC$ as the \emph{covariance matrix} of $G$.

Anticipating that this will be of some relevance later, we note that 
a positive $G$ can serve as a probability operator
(alternatively called ``state operator'' or ``density operator'').
Then Heisenberg's uncertainty relation requires
\begin{equation}
  \label{eq:HeisUR}
  \expect{a\adj a}\expect{aa\adj}\geq\expect{{a\adj}^2}\expect{a^2}
\end{equation}
so that $n$ and $m$ of a positive $G$ must be such that
\begin{equation}
  \label{eq:posG}
  n(n+1)\geq m^*m\quad\mbox{or}\quad n\geq\sqrt{m^*m+\tfrac{1}{4}}-\thalf\,.
\end{equation}

As a compact statement about the covariance matrix $\mC$, this appears as
\begin{equation}
\label{eq:pos0}
   \mC+\thalf\mE\geq0\;.
\end{equation}
Here we recall that one derivation of (\ref{eq:HeisUR}) simply exploits the
positivity of operators of the form $X\adj X$.
Choose
\begin{equation}
X=\ma\adj\mE\mx
\qquad\mbox{with}\quad\mx=\col{x\\y^*}
\end{equation}
so that
\begin{equation}
\expect{\mx\adj\mE\ma\ma\adj\mE\mx}=\mx\adj\mE\expect{\ma\ma\adj}\mE\mx\geq0
\end{equation}
has to hold for any $\mx$.
With (\ref{eq:n+m'}), the recognition that (for $G\geq0$)
\begin{equation}
\mE\expect{\ma\ma\adj}\mE=\mE\Tr{\ma\ma\adj G}\mE=\mC+\thalf\mE
\end{equation}
then establishes (\ref{eq:pos0}) as a necessary property of any positive $G$.

In (\ref{eq:CtoG}), $G$ is expanded in the Weyl basis%
\footnote{Concerning Weyl's unitary operator basis,
the seminal papers by Weyl (1927) and Schwinger (1960)
are recommended reading.\cite{Weyl27,Schwinger60}
A recent textbook account is given in chapters 1.14--1.16
of Ref.~\refcite{QM-SAM}.}
that consists of the unitary operators
$\Exp{\mz\adj\mE\ma}=\Exp{-\ma\adj\mE\mz}$.
Equivalently, we can use the Hermitian Wigner basis%
\footnote{Concerning Wigner functions,
the seminal papers by Wigner (1932) and Moyal (1949)
are recommended reading,\cite{Wigner32,Moyal49}
and so are the more recent reviews
by Tatarskii,\cite{Tatarskii83}
by Balasz and Jennings.\cite{BalJen84}
and by Hillery \textit{et al.},\cite{Hill+al84}
and also the textbook expositions by Scully and Zubairy,\cite{ScuZub97}
and Schleich.\cite{Schleich01}}
for another expansion of the same $G$.
The Wigner basis comprises the operators
\begin{equation}
  \label{eq:WigBas}
  2(-1)\power{(a\adj-z^*)(a-z)}=
\Exp{-\mz\adj\mE\ma}\,2(-1)\power{a\adj a}\,\Exp{\mz\adj\mE\ma}\,,
\end{equation}
which are obtained from the parity operator $(-1)\power{a\adj a}$
by unitary displacements;%
\footnote{Perhaps the first to note the intimate connection
between the Wigner function and the parity operator was Royer;\cite{Royer77}
in the equivalent language of the Weyl quantization scheme the analogous
observation was made a bit earlier by Grossmann.\cite{Grossmann76}
A systematic study from the viewpoint of operator bases is given
in Ref.~\refcite{bge89}. ---
The appearance of the parity operator is central to
experimental schemes for measuring Wigner functions
directly.\cite{EngStWa93,BaWod96,LutDav97}}
the factor of $2$ normalizes them to unit trace.
This gives
\begin{equation}
  \label{eq:WtoG}
  G(\ma)=\int\!(\D\mz)\,2(-1)\power{(a\adj-z^*)(a-z)}W(\mz)\,,
\end{equation}
where
\begin{eqnarray}
  W(\mz)&=&\Tr{2(-1)\power{(a\adj-z^*)(a-z)}G(\ma)}
\nonumber\\
        &=&\Tr{2(-1)\power{a\adj a}G(\ma+\mz)}
\end{eqnarray}
is the Wigner function to $G(\ma)$.
A real Wigner function, $\bigl[W(\mz)\bigr]^*=W(\mz)$,
is associated with a Hermitian operator,
$\bigl[G(\ma)\bigr]\adj=G(\ma)$.

Since Fourier transformation relates the bases to each other,
\begin{eqnarray}
&&\hspace*{-2em}
\Exp{\mz\adj\mE\ma}=\int\!(\D\mz')\,\Exp{\mz\adj\mE\mz'}
2(-1)\power{(a\adj-{z'}^*)(a-z')}\,,
\nonumber\\[2ex]
\label{eq:Weyl<-->Wig}
&&\hspace*{-2em}
2(-1)\power{(a\adj-z^*)(a-z)}=\int\!(\D\mz')\,\Exp{\mz\adj\mE\mz'}
\Exp{-\ma\adj\mE\mz'}\,,
\end{eqnarray}
the (real) Wigner function $W$ and the characteristic function $C$ are Fourier
transforms of one another,
\begin{eqnarray}
   W(\mz)&=&\int\!(\D\mz')\,\Exp{-\mz\adj\mE\mz'}C(\mz')\,,
\nonumber\\ 
   C(\mz)&=&\int\!(\D\mz')\,\Exp{-\mz\adj\mE\mz'}W(\mz')\,,
\end{eqnarray}
and, therefore, the Wigner function is also a Gaussian,
\begin{equation}
  \label{eq:Wdef}
  W(\mz)=\sqrt{\det\mW}\;\Exp{-\half\mz\adj\mW\mz}
\end{equation}
where
\begin{equation}
 \Tr{G}=1 \quad \mbox{is equivalent to} \quad \int (\D\mz)\, W(\mz)= 1\,.
\end{equation}
Note that $\bigl[C(\mz)\bigr]^*=C(-\mz)$ and
$\bigl[W(\mz)\bigr]^*=W(\mz)$ imply each other.

Perhaps the simplest verification of (\ref{eq:Weyl<-->Wig}) combines
the Baker-Hausdorff identity
\begin{eqnarray}
  \label{eq:BakHaus}
 \Exp{\mz\adj\mE\ma}&=&\Exp{-z a\adj+z^*a}
\nonumber\\
&=&\Exp{-z a\adj}\Exp{z^*a}\Exp{-\half z^*z}
=\;:\Exp{-z a\adj+z^*a}:\;\Exp{-\half z^*z}
\nonumber\\
&=&\;:\Exp{\mz\adj\mE\ma-\frac{1}{4}\mz\adj\mz}:
\end{eqnarray}
and the normally-ordered form of the displaced parity operator,
\begin{equation}
  \label{eq:normordWigBas}
\Exp{\ma\adj\mE\mz}\,2(-1)\power{a\adj a}\,\Exp{\mz\adj\mE\ma}
     =\,:2\Exp{-(\ma\adj-\mz\adj)(\ma-\mz)}:\;,
\end{equation}
with the basic Fourier-Gauss integral
\begin{equation}
  \label{eq:FourGauss}
  \int\!(\D\mz)\,\Exp{-\half\mz\adj\mA\mz}\Exp{\mz\adj\mx}
=\frac{1}{\sqrt{\det\mA}}\Exp{\half\mx\trans\mT\mA^{-1}\mx}\;,
\end{equation}
valid for all matrices $\mA=\mT\mA\trans\mT>0$ and all columns $\mx$,
whether $\mx\trans\mT$ is simply related to $\mx\adj$ or not.

Upon using (\ref{eq:BakHaus}) in (\ref{eq:CtoG})
or (\ref{eq:normordWigBas}) in (\ref{eq:WtoG})
we find the normally ordered form of $G(\ma)$,
\begin{equation}
  \label{eq:Gnormord}
  G(\ma)=\sqrt{\det\mQ}\;:\Exp{-\half\ma\adj\mQ\ma}:\;,
\end{equation}
which is another Gaussian function.
Fourier-Gauss integrals connect the various ways of writing $G(\ma)$
and, accordingly, the matrices $\mC$, $\mW$, and $\mQ$ must be simply
related. Indeed, one finds
\begin{eqnarray}
  \mC&=&\mE\mW^{-1}\mE=\mE\mQ^{-1}\mE-\thalf\mI\,,
\nonumber\\
  \mW&=&\mE\mC^{-1}\mE=\bigl(\mQ^{-1}-\thalf\mI\bigr)^{-1}\,,
\nonumber\\
  \mQ&=&\mE\bigl(\mC+\thalf\mI\bigr)^{-1}\mE
        =\bigl(\mW^{-1}+\thalf\mI\bigr)^{-1}\,,
\label{eq:CWQtoCWQ}
\end{eqnarray}
where $\mI$ is the $2\times2$ unit matrix.
The symmetry property (\ref{eq:Csymmetry}) of matrix $\mC$ is inherited by
matrices $\mW$ and $\mQ$.

Since $\mE\mC\mE$, $\mW$, $\mQ$ are functions of each other, these three
matrices commute with one another.
In fact, as long as we are dealing only with  $2\times2$ matrices,
Hermitian and with identical diagonal values, identities such as
$\mE\mC\mE=\mC^{-1}\det\mC$ can be used to achieve a further
simplification,
\begin{equation}
  \mW=\frac{\mC}{\det\mC}\,,\quad
  \mQ=\frac{\mC+\half\mI}{\det(\mC+\half\mI)}\,,
\end{equation}
but this is particular to $2\times2$ matrices and does not
hold for the $4\times4$ matrices in Sec.~\ref{sec:2D}.

In more explicit terms, we have
\begin{eqnarray}
\mW&=&\left(\Tr{\ma\ma\adj G}-\thalf\mE\right)^{-1}\,,
\nonumber\\
\mQ&=&\left(\Tr{\ma\ma\adj G}+\thalf(\mI-\mE)\right)^{-1}\,,
\end{eqnarray}
or
\begin{equation}
  \mW=\frac{1}{(n+\thalf)^2-m^*m}\matr{n+\thalf & m \\ m^* & n+\thalf}
\end{equation}
and
\begin{equation}
  \label{eq:explM}
  \mQ=\frac{1}{(n+1)^2-m^*m}\matr{n+1 & m \\ m^* & n+1}
\equiv\matr{1-\nu & \mu \\ \mu^* & 1-\nu}\,.
\end{equation}
They obey $(2\mI+\mW)(2\mI-\mQ)=4\mI$, as they should, which one verifies
easily by inspection.

It is time to note that the transitions from $\mC$ to $\mW$ and $\mQ$ are only
possible if the respective Fourier integrals are not singular, which requires
\begin{equation}
\label{eq:V>0}
\mC>0
\end{equation}
or, explicitly,
\begin{equation}
  \label{eq:exist}
  n+\thalf>\sqrt{m^*m}=|m|\,.
\end{equation}
Values of $n$ and $m$ that violate this condition will, therefore, not be
considered at all.
The determinant of $\mC$ is then positive, and so are the determinants of
$\mW$ and $\mQ$,
\begin{eqnarray}
  \det\mC&=&(n+\thalf)^2-m^*m>0\,,\nonumber\\
  \label{eq:detW}
  \det\mW&=&(\det\mC)^{-1}>0\,,\nonumber\\
  \det\mQ&=&\frac{1}{(n+1)^2-m^*m}>0\,.
\end{eqnarray}

\subsection{Positivity criteria}
Owing to its simple Gaussian form, operator $G$ must be unitarily equivalent
to the basic Gaussian $G_0$,
\begin{equation}
  \label{eq:BasGauss}
  G_0=(1-g)g\power{a\adj a}=(1-g)\;:\Exp{-(1-g)a\adj a}:\;,
\end{equation}
where $-1\leq g<1$ ensures a finite trace. In other words, we have
\begin{equation}
  \label{eq:G+G0}
G=U\adj G_0U
\end{equation}
with some unitary $U$ that effects a linear transformation on $a$ and
$a\adj$,
a \emph{squeezing transformation} in the jargon of quantum optics.
Its most general form is
\begin{eqnarray}
  U\adj aU&=&a\Exp{\I\phi}\cosh\theta+a\adj\Exp{\I\varphi}\sinh\theta\,,
\nonumber\\
  U\adj a\adj U&=&a\adj\Exp{-\I\phi}\cosh\theta+a\Exp{-\I\varphi}\sinh\theta
\end{eqnarray}
or, compactly,
\begin{equation}
  \label{eq:U2a}
  U\adj\ma U=U\adj\col{a \\ a\adj}U=\mU\col{a \\ a\adj}=\mU\ma
  \,,\quad
  U\adj\ma\adj U=\ma\adj\mU\adj
\end{equation}
with
\begin{equation}
  \label{eq:U2b}
  \mU=\matr{\Exp{\I\phi}\cosh\theta &\Exp{\I\varphi}\sinh\theta \\
            \Exp{-\I\varphi}\sinh\theta & \Exp{-\I\phi}\cosh\theta }\,,
\end{equation}
which is characterized by three real parameters: $\theta$, $\phi$, $\varphi$.
In the present context, only the relative phase $\phi-\varphi$ enters, so that
the initial parameters $n$ and $m$ determine $g$, $\theta$, and
$\phi-\varphi$.

Note that the $2\times2$ matrix $\mU$ that is thus associated with the unitary
operator $U$ is not a unitary matrix itself.
Rather it obeys
\begin{equation}
  \label{eq:mU1}
 \mU\adj\mE\mU=\mE\quad\mbox{or}\quad  \mU^{-1}=\mE\mU\adj\mE
\end{equation}
to maintain the fundamental commutation relation
\begin{equation}
aa\adj-a\adj a=1 \quad\mbox{or}\quad \ma\adj\mE\ma=-1\,,
\end{equation}
and in addition
\begin{equation}
  \label{eq:mU2}
   \mT\mU\trans=\mU\adj\mT
\end{equation}
must hold for consistency with $\ma\trans=\ma\adj\mT$.

The resulting relation between the characteristic functions of $G_0$ and $G$
amounts to
\begin{equation}
  \label{eq:mC0->mC}
  \mC=\mU\adj\mC_0\mU
\end{equation}
and, as a consequence of (\ref{eq:CWQtoCWQ}) in conjunction
with (\ref{eq:mU1}), we find
\begin{equation}
  \label{eq:mW0->mW}
  \mW=\mU\adj\mW_0\mU
\end{equation}
and
\begin{equation}
  \label{eq:mQ0->mQ}
  \mQ=\Bigl[\bigl(\mU\adj\mQ_0\mU\bigr)^{-1}
            +\thalf\mE\bigl(\mI-\mU\adj\mU\bigr)\mE\Bigr]^{-1}\;.
\end{equation}
We remark that the statement (\ref{eq:mW0->mW}) about the Wigner functions is,
of course, consistent with the general ob\-ser\-va\-tion%
\footnote{The statement is actually true for rather arbitrary
linear similarity transformations,
not just for linear unitary transformations, and it applies
to multidimensional Wigner functions.
Somewhat surprisingly, this important transformation property is not as widely
known as it should be.
Various special cases are demonstrated in
Refs.~\refcite{GarCalMosh80}, \refcite{bge89}, and \refcite{EkKni90}:
linear unitary transformations for one degree of freedom;\cite{GarCalMosh80}
linear similarity transformations (unitary or not)
for one degree of freedom;\cite{bge89}
linear unitary transformations for many degrees of freedom.\cite{EkKni90}
And Ref.~\refcite{EngFuPil02} deals with the general case of
linear similarity  transformations (unitary or not)
for many degrees of freedom.}
that linear transformations
on the operators $a,a\adj$ are reflected by exactly the same transformation
on $z,z^*$ in $W$.

Further, we note that $\mQ\neq\mU\adj\mQ_0\mU$ unless $\mU$
commutes with $\mE$,
and this is as it should be: transformations with $\mE\mU\mE\neq\mU$ turn $a$
into a linear combination of $a$ and $a\adj$, so that the meaning of normal
ordering is altered.
The extra term in (\ref{eq:mQ0->mQ}) takes just that into account.

Let us use the matrices $\mC$ and $\mC_0$ of the characteristic functions to
find the relations between the initial parameters $n,m$ and the new parameters
$g,\theta,\phi-\varphi$.
With $\mC$ of (\ref{eq:mCdef}) and
\begin{equation}
   \mC_0=\half\frac{1+g}{1-g}\mI\,,
\end{equation}
in (\ref{eq:mC0->mC}) we have
\begin{eqnarray}
  n+\half&=&\half\frac{1+g}{1-g}\cosh(2\theta)\,,
\nonumber\\
  m&=&\half\frac{1+g}{1-g}\Exp{-\I(\phi-\varphi)}\sinh(2\theta)\,,
  \label{eq:Ueff}
\end{eqnarray}
and, in particular,
\begin{equation}
 (n+\thalf)^2-m^*m=\frac{1}{4}\left(\frac{1+g}{1-g}\right)^2\,.
\end{equation}
Thus $g$ is given by
\begin{equation}
  \label{eq:expl_g}
  g=\frac{\bigl[(n+\half)^2-m^*m\bigr]^{\half}-\half}
         {\bigl[(n+\half)^2-m^*m\bigr]^{\half}+\half}
\,,
\end{equation}
and since the eigenvalues of $G$ are $(1-g)g^k$ with $k=0,1,2,\ldots$ we find
that $G\geq0$ requires $g\geq0$, which in turn says that the argument of the
square root must be at least $\frac{1}{4}$, and this is precisely the
constraint (\ref{eq:posG}).
In other words: Condition (\ref{eq:pos0}) is both necessary and sufficient for
the positivity of $G$.

These considerations are instructive, but they are not really needed if we
just want to find the eigenvalues of $G$, for which purpose knowledge of
$\theta$ and $\phi-\varphi$ is obsolete.
A direct method proceeds from the observation that
\begin{equation}
  \Tr{G^2}=\sum_{k=0}^{\infty}\left[(1-g)g^k\right]^2=\frac{1-g}{1+g}
\end{equation}
and employs the Wigner function for calculating this trace,
\begin{equation}
  \label{eq:trGG'}
  \Tr{G^2}=\int\!(\D\mz)\,\left[W(\mz)\right]^2
   =\half\sqrt{\det\mW}\,.
\end{equation}
In conjunction with (\ref{eq:detW}) this reproduces (\ref{eq:expl_g}).

Note that condition (\ref{eq:trGG'}) follows from the general
property of a density operator for which
\begin{equation}
\label{W21Da}
 \Tr{G^2}\leq1 \quad \mbox{is equivalent to} 
\quad \int (\D\mz)\, W(\mz)^2\leq 1\,.
\end{equation}
The discussion above tells us that condition
(\ref{W21Da}), which leads to
\begin{equation}
  \label{W21Db}
\sqrt{\det\mW} \leq 2\,,
\end{equation}
is both necessary and sufficient for the positivity of $G$ in
one dimension. 
As we shall see in Sec.~\ref{sec:2D}, however, this condition
is not strong enough to guarantee the positivity of
$G$ in two and more dimensions.

We do not even need to know the eigenvalues of $G$ if
checking $G\geq0$ is all that we are interested in. For, the normally
ordered form (\ref{eq:Gnormord}) reads more explicitly
\begin{equation}
  \label{eq:Gnormord'}
  G=\sqrt{\det\mQ}\,\Exp{-\half\mu{a\adj}^2}:\Exp{-(1-\nu)a\adj a}:\;
                  \Exp{-\half\mu^*a^2}\,.
\end{equation}
This has the structure $G=S\adj \tG S$ with some $S$ and
\begin{equation}
     \tG=(1-\nu)\;:\Exp{-(1-\nu)a\adj a}:\;=(1-\nu)\nu\power{a\adj a}\,,
\end{equation}
or
\begin{equation}
  \label{eq:tG+tM}
  \tG=\sqrt{\det\tmQ}\;:\Exp{-\half\ma\adj\tmQ\ma}:
\end{equation}
with
\begin{equation}
\tmQ=\matr{1-\nu&0\\0&1-\nu}=\half\bigl(\mQ+\mE\mQ\mE\bigr)\,.
\end{equation}
Consequently, $G\geq0$ is ensured by $\tG\geq0$,
and this just requires $\nu\geq0$,
or
\begin{equation}
\label{eq:tmQ<1}
\mI-\tmQ\geq0\,.
\end{equation}
Now, upon recalling how $\nu$ is related to $n$ and $m$ in (\ref{eq:explM}),
\begin{equation}
  \label{eq:nu}
  \nu=\frac{n(n+1)-m^*m}{(n+1)^2-m^*m}\,,
\end{equation}
we find, once more, that (\ref{eq:pos0}) is the positivity criterion.

Having found the value of $g$, the unitary transformation of (\ref{eq:G+G0})
can be identified.
For this purpose we return to (\ref{eq:mC0->mC}) and write it in the
equivalent forms
\begin{equation}
  \mC\mE\mU\adj=\mU\adj\mC_0\mE\qquad\mbox{and}\qquad\mU\mE\mC=\mE\mC_0\mU\,,
\end{equation}
where
\begin{equation}
\mE\mC_0=\half\frac{1+g}{1-g}\matr{1&0\\0&-1}=\mC_0\mE\,.
\end{equation}
Therefore,
the eigenvalues of the $2\times2$ matrices
$\mE\mC$ and $\mC\mE$ are $\pm\half(1+g)/(1-g)$,
the columns of $\mU\adj$ are the respective eigencolumns of $\mC\mE$,
and the rows of $\mU$ are the eigenrows of $\mE\mC$.
With (\ref{eq:Ueff}) and (\ref{eq:expl_g}), it is a matter of inspection
to verify these statements for $\mC$ of (\ref{eq:mCdef})
and $\mU$ of (\ref{eq:U2b}).

\subsection{P-representable positive Gaussian operators}
\label{sec:1D-Prepr}
For $g=0$, the basic Gaussian $G_0$ of (\ref{eq:BasGauss}) is
$:\Exp{-a\adj a}:$, the projector to the oscillator ground state.
Unitary displacements turn it into $:\Exp{-(a\adj-z^*)(a-z)}:$, which
project onto the coherent states, the eigenbras of $a\adj$ and
eigenkets of $a$ with respective eigenvalues $z^*$ and $z$.

A positive Gaussian operator, $G>0$, is said to be P-representable if one can
write it as a mixture of coherent states,
\begin{equation}
  \label{eq:P.1}
  G=\int\!(\D\mz)\;:\Exp{-(a\adj-z^*)(a-z)}:\;
    \sqrt{\det\mP}\,\Exp{-\half\mz\adj\mP\mz}
\end{equation}
with $\mP>0$. The limiting case of $G=:\Exp{-a\adj a}:$, when%
\footnote{See (\ref{eq:delta-def}) below for the definition of the
two-dimensional Dirac delta function.}
\begin{equation}
  \sqrt{\det\mP}\,\Exp{-\half\mz\adj\mP\mz}\to\delta(\mz)\,,
\end{equation}
so that $\mP\to\infty$ in some sense, need not concern us too much.
For the sake of notational simplicity, we exclude it by considering only $G>0$,
rather than $G\geq0$.

For a P-representable Gaussian, we have
\begin{eqnarray}
  \label{eq:P.3a}
  \mQ&=&\bigl(\mP^{-1}+\mI\bigr)^{-1}\,,\nonumber\\
  \mW&=&\bigl(\mP^{-1}+\thalf\mI\bigr)^{-1}\,,\nonumber\\
  \mC&=&\mE\mP^{-1}\mE+\thalf\mI\,,
\end{eqnarray}
and
\begin{equation}
  \label{eq:P.4}
  \mP=\mE\bigl(\mC-\thalf\mI\bigr)^{-1}\mE
     =\bigl(\mW^{-1}-\thalf\mI\bigr)^{-1}
     =\bigl(\mQ^{-1}-\mI\bigr)^{-1}\,.
\end{equation}
So, a given $G$ is P-representable if
\begin{equation}
  \mC-\thalf\mI>0\,,\quad\mW^{-1}-\thalf\mI>0\,,\quad\mQ^{-1}-\mI>0\,,
\end{equation}
which requires
\begin{equation}
  \label{eq:Prep1}
  n>\bigl|m\bigr|\,.
\end{equation}
There are, therefore, positive Gaussians operators that are not
P-re\-pre\-sent\-able, namely those with
\begin{equation}
  \bigl|m\bigr|>n>\sqrt{m^*m+\tfrac{1}{4}}-\thalf\,.
\end{equation}

All positive Gaussian operators are, however, unitarily equivalent to a
P-re\-pre\-sent\-able one,
because $G_0$ of (\ref{eq:BasGauss}) is P-representable if $g>0$, as
\begin{equation}
\label{eq:P.8}
  (1-g)g\power{a\adj a}=\int\!(\D\mz)\;:\Exp{-(a\adj-z^*)(a-z)}:
\ \bigl(g^{-1}-1\bigr)\Exp{-(g^{-1}-1)z^*z}
\end{equation}
shows explicitly.
If $G>0$ is P-representable, then the unitary transformation of
(\ref{eq:BasGauss}) amounts to
\begin{equation}
  \label{eq:P.9}
  \mP=\Bigl[\bigl(\mU\adj\mP_0\mU\bigr)^{-1}
            -\thalf\mE\bigl(\mI-\mU\adj\mU\bigr)\mE\Bigr]^{-1}\;,
\end{equation}
where we see an extra term, similar to the one in (\ref{eq:mQ0->mQ}),
which reflects the injunction of normal ordering that is
inherent in (\ref{eq:P.1}).

The unitary transformation of (\ref{eq:G+G0}) that relates the given positive
$G$ to the special P-representable $G_0$ of (\ref{eq:BasGauss}) and
(\ref{eq:P.8}), is just one transformation of many, which are all such that
$UGU\adj$ is a P-representable Gaussian.
In terms of the respective characteristic functions, these transformations
are such that
\begin{equation}
  \label{eq:P.10}
   \mC=\matr{n+\half & m \\ m^* &n+\half} \rightarrow
  {\mU\adj}^{-1}\mC\mU^{-1}=
  \mE\mU\mE\mC\mE\mU\adj\mE
  =\matr{N+\half & M \\ M^* & N+\half}
\end{equation}
with
\begin{equation}
  \label{eq:Prep2}
  N >\bigl|M\bigr| \,,
\end{equation}
where $\mU$ is of the form (\ref{eq:U2b}) and obeys (\ref{eq:mU1}).
For example, with $\Real\bigl(m\Exp{\I\phi-\I\varphi}\bigr)=|m|$ ensured by
choosing $\phi$ and $\varphi$ accordingly, we get
\begin{eqnarray}
\label{eq:NMdef}
 N+\thalf &=&(n+\thalf) \cosh (2\theta)-|m|\sinh(2\theta)\,, \nonumber\\
  M & = & \Exp{\I\phi+\I\varphi}\bigl[|m|\cosh
 (2\theta)-(n+\thalf)\sinh(2\theta)\bigr]\,,
\end{eqnarray}
and then (\ref{eq:Prep2}), the condition that $UGU\adj$ is a
P-representable Gaussian, amounts to
\begin{equation}
\label{eq:boundtheta}
 \frac{1}{2n+1 -2\bigl|m\bigr|}\leq
\Exp{2\theta} \leq 2n+1 +2\bigl|m\bigr| \,.
\end{equation}
Owing to (\ref{eq:exist}), the lower bound is assuredly positive, and since
(\ref{eq:posG}) holds for a positive $G$, the upper bound is certainly larger
than the lower one so that the range for $\Exp{2\theta}$ is not empty.

Any $\theta$ from this range will serve the purpose of relating the given $G$
to a P-representable one.
In particular we note that $\theta=0$ is permissible (of course)
if $G$ itself is P-representable ($n>|m|$).
The special transformation of (\ref{eq:G+G0}), for which $UGU\adj=G_0$ and
thus $M=0$, obtains when $\Exp{2\theta}$ is the geometric mean of the bounds,
so that
\begin{equation}
 \Exp{4\theta_{0}} =\frac{n+\half
+\bigl|m\bigr|}{n+\half -\bigl|m\bigr|}
\end{equation}
identifies the $\theta$ value for this distinguished squeezing transformation.

\subsection{Transposed Gaussians}
As a preparation for a later discussion in the context of Gaussian states of
two entangled harmonic oscillators, let us briefly discuss what happens when
an operator transposition is done.
First of all, we must note that transposing an operator
is a representation-dependent operation.
For a chosen basis of Hilbert state vectors, $\ket{k}$, the
transpose $G\trans$ of $G$ is defined by
\begin{equation}
  \bra{k}G\trans\ket{k'}=\bra{k'}G\ket{k}\;.
\end{equation}
Clearly, if the $\ket{k}$'s are eigenkets of $G$, there is no difference
between $G$ and $G\trans$.

We shall consider two procedures based on the position and momentum
representations associated with the eigenstates of
$q=2^{-\half}(a\adj+a)$ and $p=2^{-\half}(\I a\adj-\I a)$, respectively.
Transposition in the $q$-representation has this effect on products of
a $q$-function and a $p$-function,
\begin{equation}
  \left[f_1(q)f_2(p)\right]\trans=f_2(-p)f_1(q)\,,
\end{equation}
and in the $p$-representation one gets
\begin{equation}
  \left[f_1(q)f_2(p)\right]\trans=f_2(p)f_1(-q)\,.
\end{equation}
We find the resulting transformation of $\Exp{z a\adj-z^*a}$ by
first noting that
\begin{eqnarray}
  \Exp{z a\adj-z^*a}
  &=&\Exp{2^{-\half}(z-z^*)q-2^{-\half}\I(z+z^*)p}
\nonumber\\
  &=&\Exp{2^{-\half}(z-z^*)q}\Exp{-2^{-\half}\I(z+z^*)p}
\Exp{-\frac{1}{4}(z-z^*)(z+z^*)}\,,
\end{eqnarray}
so that
\begin{eqnarray}
  \left[\Exp{z a\adj-z^*a}\right]\trans&=&
\Exp{2^{-\half}\I(z+z^*)p}
\Exp{2^{-\half}(z-z^*)q}
\Exp{-\frac{1}{4}(z-z^*)(z+z^*)}\nonumber\\[-2ex]
&=&
\Exp{2^{-\half}(z-z^*)q+2^{-\half}\I(z+z^*)p}
\nonumber\\
&=&\Exp{z a-z^*a\adj}
\end{eqnarray}
in the $q$-representation, whereas
\begin{eqnarray}
  \left[\Exp{z a\adj-z^*a}\right]\trans&=&
\Exp{-2^{-\half}\I(z+z^*)p}
\Exp{-2^{-\half}(z-z^*)q}
\Exp{-\frac{1}{4}(z-z^*)(z+z^*)}\nonumber\\[-2ex]
&=&
\Exp{-2^{-\half}(z-z^*)q-2^{-\half}\I(z+z^*)p}
\nonumber\\
&=&\Exp{z^*a\adj-za}
\end{eqnarray}
in the $p$-representation. These are compactly summarized in
\begin{eqnarray}
\hspace*{-2em}\left[\Exp{\ma\adj\mE\mz}\right]\trans
=\left[\Exp{-\mz\adj\mE\ma}\right]\trans
&=&\Exp{\pm\ma\adj\mT\mE\mz}=\Exp{\mp\ma\adj\mE\mT\mz}
\nonumber\\
&=&\Exp{\mp\mz\adj\mT\mE\ma}=\Exp{\pm\mz\adj\mE\mT\ma}
\end{eqnarray}
where the upper sign refers to the $q$-representation, and the
lower sign to the $p$-representation.
Not surprisingly, we encounter the transposition
matrix $\mT$ of (\ref{eq:mTdef}).

In $C$ of (\ref{eq:Cdef}) and $W$ of (\ref{eq:Wdef}) we thus have
\begin{equation}
\mz\to\pm\mT\mz\,,\qquad\mz\adj\to\pm\mz\adj\mT\,,
\end{equation}
and
\begin{equation}
\ma\to\mp\mT\ma\,,\qquad\ma\adj\to\mp\ma\adj\mT
\end{equation}
apply in the normally ordered form of (\ref{eq:Gnormord}).
Since the sign is irrelevant for the Gaussians that we are concerned
with, we have in both cases
\begin{equation}
  \label{eq:trans-mC,mW,mQ}
  \mC\to\mT\mC\mT\,,\quad
  \mW\to\mT\mW\mT\,,\quad
  \mQ\to\mT\mQ\mT\,,\quad
  \mP\to\mT\mP\mT\,,
\end{equation}
which are, of course, consistent with (\ref{eq:CWQtoCWQ})
as well as (\ref{eq:P.3a}) and (\ref{eq:P.4}).

Indeed, the net effect is simply $m\leftrightarrow m^*$,
and therefore $G\trans$ has the same eigenvalues as $G$, so that $G$
and $G\trans$ are unitarily equivalent.
The situation is markedly different when a \emph{partial} transposition is
done on a Gaussian state of two harmonic oscillators.

\subsection{Examples}
We conclude this section on one-dimensional Gaussian operators 
by giving two explicit examples.

\subsubsection{Parity Operator}

The first example is the parity operator that is used to form the
Wigner basis. We have previously noted that the transition
from $\mC$ to $\mW$ and $\mQ$ is nonsingular if $\mC >0$.
Actually, the limit $\mC\geq0$, $C\not>0$ can be included with a bit of
caution,
and $\mC=0$ is actually needed in the construction of the
Wigner representation of the Gaussian operators (\ref{eq:WtoG}).
For this choice of $\mC$, we obtain: $\mQ =\thalf\,\mI$, and
accordingly the corresponding Wigner function of such an operator
is given by a singular expression that can be normalized:
\begin{equation}\label{eq:WParity}
    W(\mz)= \delta(\mz)\mbox{\qquad with\qquad} \int\!(\D\mz)\, W(\mz)=1\,,
\end{equation}
where
\begin{equation}
  \label{eq:delta-def}
  \delta(\mz)\equiv 2\pi\delta(q')\delta(p')\,,
  \qquad \int\!(\D\mz)\, \delta(\mz)f(\mz)=f\biggl(\mz=\col{0\\0}\biggr)
\end{equation}
for both parameterizations (\ref{eq:PhSpVariables}) and (\ref{eq:dz2}).
Using this relation, we obtain from (\ref{eq:WtoG}) that the
resulting Gaussian operator
\begin{equation}\label{eq:parity}
G= 2 \;:\Exp{- 2a\adj a}:\ \ = 2\;(-1)\power{a\adj a}
\end{equation}
is twice the well known parity operator. This
operator is non-positive and normalized to unit trace, $\Tr{G}=1$. 
Note that the Wigner function (\ref{eq:WParity}) is singular, normalized and
positive, while the corresponding Gaussian operator is normalized but not 
positive. 
Upon unitarily shifting the parity operator by complex numbers $z$, 
we reproduce all elements of the Wigner basis (\ref{eq:WigBas}).

\subsubsection{Pure Gaussian State}

The second example is a Gaussian operator describing a pure
quantum state. It follows from (\ref{eq:trGG'})
that a Gaussian state is pure if
\begin{equation}
 \det\mC = (\det\mW)^{-1} =(n+\thalf)^2-m^*m=\frac{1}{4}
\end{equation}
holds, which amounts to
\begin{equation}\label{eq:pure2}
|m| = \sqrt{n(n+1)}\,.
\end{equation}
In this case the general formula (\ref{eq:Gnormord'}) reduces to a
projector
\begin{equation}
  \label{eq:purestate}
  G= \proj{\mu}=\sqrt{\det\mQ}\,\Exp{-\half\mu{a\adj}^2}\proj{0} \;
                  \Exp{-\half\mu^*a^2}\,,
\end{equation}
where
\begin{equation}
|\mu| =\sqrt{\frac{n}{n+1}}\,,\qquad
\det\mQ = \frac{1}{n+1}=1-\mu^*\mu\,,
\end{equation}
and
\begin{equation}
  \label{eq:vacproj}
  \proj{0}= \,:\Exp{-a\adj a}:
\end{equation}
projects on the $0$th Fock state, the ground state of the
harmonic oscillator: $a\ket{0}=0$.

In (\ref{eq:pure2}) we recognize the border case of (\ref{eq:posG}), 
and thus of (\ref{eq:HeisUR}).
This tells us that one can check the purity of a Gaussian sate of a 
one-dimensional oscillator by measuring the expectation values of 
$a\adj a$ and $a^2$, and then verifying that the equal sign holds in    
(\ref{eq:HeisUR}).

From (\ref{eq:purestate})--(\ref{eq:vacproj}) we see that the ket vector 
of the pure Gaussian state has the form
\begin{equation}
\label{eq:psi} \ket{\mu} =
(1-\mu^*\mu)\power{\frac{1}{4}}\,
\Exp{-\half\mu{a\adj}^2}\ket{0}\,.
\end{equation}
Now assuming, for  simplicity, that  $\mu$ is real and
positive, $\mu=|\mu|>0$,
we recognize in this expression a squeezed state of a
one-dimensional harmonic oscillator, whose  position wave function is
\begin{equation}
\label{eq:psiq} \psi_{\mu}(q) = \braket{q}{\mu}=
\left(\frac{\kappa}{\pi}\right)\power{\frac{1}{4}}
\Exp{-\frac{1}{2}\kappa q^{2}}
\end{equation}
with
\begin{equation}
 \kappa=\frac{1+\mu}{1-\mu}=\frac{\sqrt{n+1}+\sqrt{n}}{\sqrt{n+1}-\sqrt{n}}\,.
\end{equation}
For $n=0$, we have $\mu=0$, $\kappa=1$; the wave function is
then that of the ground state of the one-dimensional
harmonic oscillator, as it should be. 

The general pure squeezed state is never P-representable because
the relation (\ref{eq:Prep1}) is never satisfied when (\ref{eq:pure2}) holds.
But as we have seen above, it can be unitarily transformed into the projector
$\proj{0}$ which is a P-representable Gaussian operator.


\section{Gaussian States of a Two-Dimensional Oscillator}\label{sec:2D}

\subsection{Parameterizations}
Things look much the same except that the number of variables doubles.
Thus now we have $a_1\adj,a_1$ and $z^*_1,z_1$ for the first degree of
freedom, as well as $a_2\adj,a_2$ and $z^*_2,z_2$ for the second, so that
\begin{equation}
  \ma=\col{a_1 \\ a_1\adj \\a_2 \\a_2\adj}\,,\quad
  \ma\adj=\row{a_1\adj,a_1,a_2\adj,a_2}\,,\qquad
  \mz=\col{z_1 \\ z^*_1 \\z_2 \\z^*_2}\,,\quad
  \mz\adj=\row{z^*_1,z_1,z^*_2,z_2}
\end{equation}
are 4-component columns and rows, and $\mC$, $\mW$, $\mQ$, $\mP$ are
$4\times4$-matrices. The number of integration variables doubles as well, of
course, and
\begin{equation}
  \int\!(\D\mz)\ldots=\int\!(\D\mz_1)\int\!(\D\mz_2)\ldots
\end{equation}
specifies how phase space integrals are to be understood.
Relations (\ref{eq:CWQtoCWQ}),
(\ref{eq:P.3a}), and (\ref{eq:P.4}) remain valid with
\begin{equation}
  \mE=\Matr{1&0&0&0 \\ 0&-1&0&0 \\ 0&0&1&0 \\ 0&0&0&-1}
\end{equation}
replacing the $2\times2$ version of $\mE$ in (\ref{eq:defE}).

Linear unitary transformations of the form (\ref{eq:G+G0}) now involve a
$4\times4$ matrix $\mU$ for the transformation of $\ma$ and $\ma\adj$ as in
(\ref{eq:U2a}), which continues to obey (\ref{eq:mU1}) and (\ref{eq:mU2})
with
\begin{equation}
   \mT=\Matr{0&1&0&0\\1&0&0&0\\0&0&0&1\\0&0&1&0}
\end{equation}
now.
With these replacements, the $4\times4$ matrices $\mC$, $\mW$, $\mQ$, and $\mP$
transform according to (\ref{eq:mC0->mC}), (\ref{eq:mW0->mW}),
(\ref{eq:mQ0->mQ}), and (\ref{eq:P.9}), respectively.

The explicit parameterization of the matrix appearing in the Gaussian
characteristic function is
\begin{equation}
  \label{eq:2DdefC}
  \mC=\Matr{n_1+\half & m_1 &  m_s & m_c  \\
            m_1^*  & n_1+\half  &  m_c^* & m_s^* \\
            m_s^* & m_c & n_2+\half & m_2 \\
            m_c^* & m_s & m_2^* & n_2+\half }
\end{equation}
and $\mC>0$ is taken for granted again.
Here, too, expanding
\begin{equation}
   \Exp{-\half\mz\adj\mC\mz}=\Tr{\Exp{\ma\adj\mE\mz}G}
  =\Tr{\Exp{z_1\phstar a_1\adj-z_1^*a_1\phadj}
      \Exp{z_2\phstar a_2\adj-z_2^*a_2\phadj}G}
\end{equation}
in powers of $\mz$ reveals the physical significance of
the numerical parameters in $\mC$, viz.\
\begin{equation}
  \begin{array}[b]{rl@{\qquad}rl}
  n_1&=\Tr{a_1\adj a_1\phadj G}\,,& m_1&=-\Tr{a_1^2G} \,,\\
  n_2&=\Tr{a_2\adj a_2\phadj G}\,,& m_2&=-\Tr{a_2^2G} \,,\\
  m_s&=\Tr{a_1\phadj a_2\adj G}\,,& m_c&=-\Tr{a_1\phadj a_2\phadj G}\,,
  \end{array}
\end{equation}
and positive two-dimensional Gaussians must obey (\ref{eq:pos0}) for
analogous reasons as in the one-dimensional case of Sec.~\ref{sec:1D}.

\emph{Partial} transposition,
in the $z^*_1,z_1\phstar$ sector only, turns
$G$ into $G\ptrans$. For $\mC$, $\mW$, $\mQ$, and $\mP$ the transition
is as given in
(\ref{eq:trans-mC,mW,mQ}) with $\mT$ replaced by
\begin{equation}
  \mT_1=\Matr{0&1&0&0 \\ 1&0&0&0 \\ 0&0&1&0 \\ 0&0&0&1}\;.
\end{equation}
We note that
\begin{equation}
  \mC\to\mE\mC\mE\qquad\mbox{amounts to}\qquad
  \left\{\begin{array}{rl}
   m_1&\to-m_1  \\ m_2&\to-m_2 \\ m_c&\to-m_c
  \end{array}\right\}
\end{equation}
with $n_1,n_2,m_s$ unaffected,
and
\begin{equation}
  \label{eq:C->TCT}
  \mC\to\mT_1\mC\mT_1\qquad\mbox{amounts to}\qquad
  \left\{\begin{array}{rl}
   m_1&\to m_1^*  \\ m_s&\to m_c^*\\ m_c&\to m_s^*
  \end{array}\right\}
\end{equation}
with $n_1,n_2,m_2$ unaffected;
analogous statements apply to
$\mQ\to\mE\mQ\mE$, $\mQ\to\mT_1\mQ\mT_1$, \emph{et~cetera}.

\subsection{Positivity criteria}
A first positivity check is the one that exploits that
$G$ must be unitarily equivalent to a (product of) basic Gaussian(s),
\begin{equation}
  \label{eq:2Dbasic}
  G=U\adj\, (1-g_1)g_1\power{a_1\adj a_1\phadj}\,
            (1-g_2)g_2\power{a_2\adj a_2\phadj} \,U
\end{equation}
with some appropriate unitary $U$. The positivity of $G$ can be tested by
checking whether $g_1$ and $g_2$ are positive, for which it suffices to see if
the smaller one of the two is positive.%
\footnote{In marked contrast to the one-dimensional case,
condition (\ref{W21Da}) alone does not guarantee
the positivity of a two-dimensional Gaussian operator. 
As a counter example put $g_1=\frac{1}{3}$, $g_2=-\frac{1}{3}$ in  
(\ref{eq:2Dbasic}), so that the resulting Gaussian has negative eigenvalues
while $\Tr{G}=\Tr{G^2}=1$.
We are reminded here that the positivity of the statistical operator for a
two-party system  requires in particular the positivity of the
reduced density operators, here characterized respectively by $g_1$ and
$g_2$.
In two dimensions, we also need to consider $\Tr{G^4}$ when verifying 
the positivity of $G$.} 

Now, information about $g_1$ and $g_2$ is available in the traces
\begin{eqnarray}
 \Tr{G^2}&=&\frac{1-g_1}{1+g_1} \frac{1-g_2}{1+g_2}\,,\quad
\nonumber\\
 \Tr{G^4}&=&\frac{(1-g_1)^4}{1-g_1^4} \frac{(1-g_2)^4}{1-g_2^4}\,.
\end{eqnarray}
The trace of $G^2$ can be expressed in terms of the $\mC$ matrix,
\begin{equation}
  \label{eq:trGG''}
  \Tr{G^2}=\frac{1}{\sqrt{\det(2\mC)}}=\frac{1}{4\sqrt{\det\mC}}\,,
\end{equation}
and upon introducing
\begin{equation}
  \barr{G}\equiv\frac{G^2}{\Tr{G^2}}=
    U\adj\, (1-g_1^2)g_1\power{2a_1\adj a_1\phadj}\,
            (1-g_2^2)g_2\power{2a_2\adj a_2\phadj} \,U\,,
\end{equation}
with matrix $\barr{\mC}$ in its characteristic function
(see Sec.~\ref{sec:sqGauss} below), we have
\begin{equation}
  \frac{\Tr{G^4}}{(\Tr{G^2})^2}=
  \Tr{\barr{G}^2}=\frac{1}{4\sqrt{\det\barr{\mC}\,}}\,.
\end{equation}
 
So, we extract the necessary information about $g_1$ and $g_2$ out of%
\,\footnote{Once $g_1$ and $g_2$ are available, a variety of unitarily
  invariant quantities can be computed, such as the von Neumann entropy of
  the mixed state represented by $G$.\cite{2Dentropy}}
\begin{equation}
4\sqrt{\det\mC}=\frac{1+g_1}{1-g_1} \frac{1+g_2}{1-g_2}
\quad \mbox{and}\quad
4\sqrt{\det\barr{\mC}}=\frac{1+g_1^2}{1-g_1^2} \frac{1+g_2^2}{1-g_2^2}\,.
\end{equation}
A straightforward, yet somewhat tedious, calculation then establishes that
${G\geq0}$ holds if both inequalities in
\begin{equation}
  \label{eq:2DposG.1}
  \tfrac{1}{16}+3\det\mC\geq4\sqrt{\det\mC}\sqrt{\det\barr{\mC}}
\leq\tfrac{1}{8}+2\det\mC
\end{equation}
are obeyed, and only then.
In deriving these two inequalities, it is taken for granted that
\begin{equation}
  -1<g_1,g_2<1\,,
\end{equation}
which is not an actual limitation because $G\adj=G$ and $\Tr{G^2}<\infty$ 
restrict the values of $g_1$ and $g_2$ to this range.

The right inequality in (\ref{eq:2DposG.1}) amounts to
\begin{equation}
  g_1g_2\geq0
\end{equation}
and the left to
\begin{equation}
  (g_1+g_2)(1+g_1)(1+g_2)\geq(g_1-g_2)^2\,.
\end{equation}
The first requires that both $g$'s are positive or both negative, but the
second excludes the possibilities $g_1\leq0,g_2<0$ and $g_1<0,g_2\leq0$.
So, taken together, the two inequalities of (\ref{eq:2DposG.1}) imply
$g_1,g_2\geq0$, that is: $G\geq0$, indeed.

Alternatively, for a second positivity check
we can apply the reasoning of (\ref{eq:Gnormord'})--(\ref{eq:nu}) to
\begin{equation}\label{eq:2DmQ}
\mQ=\Matr{1-\nu_1 & \mu_1 & \mu_s & \mu_c \\
          \mu_1^* & 1-\nu_1 & \mu_c^* & \mu_s^* \\
          \mu_s^* & \mu_c & 1-\nu_2 & \mu_2 \\
          \mu_c^* & \mu_s & \mu_2^* & 1-\nu_2}\;,
\end{equation}
where now
\begin{equation}
\tmQ=\Matr{1-\nu_1 & 0 & \mu_s & 0 \\
            0 & 1-\nu_1 & 0 & \mu_s^* \\
           \mu_s^* & 0 & 1-\nu_2 & 0 \\
            0 & \mu_s & 0 & 1-\nu_2}=\half(\mQ+\mE\mQ\mE)
\end{equation}
shows up in the $4\times4$ version of (\ref{eq:tG+tM}).
Since $\tmQ$ commutes with $\mE$, we only need to consider
unitary transformations
with $\mU\mE=\mE\mU$ and thus $\mU\adj\mU=\mI$ when diagonalizing $\tG$,
so that the complication of the extra term in (\ref{eq:mQ0->mQ}) is of
no concern here.
Therefore, the criterion (\ref{eq:tmQ<1}) applies
to the two-dimensional case as well, and we are asked
to check if the eigenvalues of $\tmQ$ exceed unity or not.
These eigenvalues are
\begin{equation}
1-\thalf(\nu_1+\nu_2)
\pm\sqrt{\bigl[\thalf(\nu_1-\nu_2)\bigr]^2+\mu_s^*\mu_s\phstar}\,,
\end{equation}
so that $G\geq0$ if
\begin{equation}
\label{eq:2Dpos,tmQ'}
\nu_1+\nu_2\geq\sqrt{(\nu_1-\nu_2)^2+4\mu_s^*\mu_s\phstar}
\end{equation}
or
\begin{equation}
\label{eq:2Dpos,tmQ}
\nu_1+\nu_2\geq0\enskip\mbox{and}\enskip\nu_1\nu_2\geq\mu_s^*\mu_s\phstar
\end{equation}
and only then.

\subsection{Squared Gaussians}\label{sec:sqGauss}
The positivity criterion (\ref{eq:2DposG.1}) needs knowledge of $\barr{\mC}$,
the $4\times4$-matrix appearing in the characteristic function of $\barr{G}$.
It is found by squaring
\begin{equation}
  G=\int\!(\D\mz)\,\Exp{\mz\adj\mE\ma-\half\mz\adj\mC\mz}\,,
\end{equation}
recalling the Baker-Hausdorff relation et cetera, with the outcome
\begin{equation}
  \barr{\mC}=\thalf\mC+\tfrac{1}{8}\mE\mC^{-1}\mE
            =\thalf\mC+\tfrac{1}{8}\mW\,.
\end{equation}
As a consequence, then
\begin{equation}
  \det\barr{\mC}
  =\tfrac{1}{16}\det\left(\mC+\tfrac{1}{4}\mE\mC^{-1}\mE\right)\,,
\end{equation}
and (\ref{eq:2DposG.1}) reads more explicitly
\begin{equation}
\tfrac{1}{16}+3\det\mC
\geq\sqrt{(\det\mC)\det\left(\mC+\tfrac{1}{4}\mE\mC^{-1}\mE\right)}
\leq\tfrac{1}{8}+2\det\mC\,.
\end{equation}

\subsection{Separable and non-separable Gaussians}\label{sec:2D-SepCrit}
As a rule, the statistical operator
$\rho^{(1\&2)}(\ma)=\rho^{(1\&2)}(\ma_1,\ma_2)$
of a two-dimensional oscillator, be it of
Gaussian shape or not, will be different from the product of the reduced
statistical operators $\rho^{(1)}(\ma_1)$, $\rho^{(2)}(\ma_2)$
that one obtains by partial tracing,
\begin{equation}
 \rho^{(1\&2)}(\ma)\neq\rho^{(1)}(\ma_1)\rho^{(2)}(\ma_2)
\end{equation}
with
\begin{equation}
\rho^{(1)}(\ma_1)=\Tr[2]{\rho^{(1\&2)}(\ma)}\,,\quad
\rho^{(2)}(\ma_2)=\Tr[1]{\rho^{(1\&2)}(\ma)}\,.
\end{equation}
But, as we mentioned in the Introduction, 
it could happen quite easily that $\rho^{(1\&2)}$ is the convex sum of such
products, in which case it represents a \emph{separable} state,
\begin{equation}
 \rho^{(1\&2)}\mbox{\ is separable if}\quad
\rho^{(1\&2)}=\sum_kw_k \rho_k^{(1)}\rho_k^{(2)}\quad\mbox{with\ }w_k>0\,.
\end{equation}
Whereas it can be rather difficult to decide if a given $\rho^{(1\&2)}$ is
separable, matters are remarkably simple for Gaussian states.

Positive Gaussians that are separable must have a positive
partial transpose --- this is Peres's necessary criterion.\cite{Peres96}
In the case of Gaussian states,
it is also sufficient.\cite{Duan+al00,Simon00,we02}

Concerning the positivity of $G\ptrans$,
please note that $\det(\mT_1\mC\mT_1)=\det\mC$ and that
\begin{eqnarray}
  \barr{\;\mT_1\mC\mT_1}&=&\thalf\mT_1\left(\mC
                   +\tfrac{1}{4}\mT_1\mE\mT_1\mC^{-1}\mT_1\mE\mT_1\right)\mT_1
\nonumber\\
                  &=&\thalf\mT_1\left(\mC
                   +\tfrac{1}{4}\mE\ptrans\mC^{-1}\mE\ptrans\right)\mT_1
\end{eqnarray}
so that a non-negative partial transpose, $G\ptrans\geq0$, is available only if
\begin{equation}
\tfrac{1}{16}+3\det\mC
\geq\sqrt{(\det\mC)\det\left(\mC
                      +\tfrac{1}{4}\mE\ptrans\mC^{-1}\mE\ptrans\right)}
\leq\tfrac{1}{8}+2\det\mC
\end{equation}
hold with
\begin{equation}
  \mE\ptrans=\mT_1\mE\mT_1=\Matr{-1&0&0&0 \\ 0&1&0&0 \\ 0&0&1&0 \\ 0&0&0&-1}\,.
\end{equation}

Probably simpler is the positivity criterion of
(\ref{eq:2DmQ})--(\ref{eq:2Dpos,tmQ}).
When applied to $G\ptrans$ it asks whether the eigenvalues of
\begin{equation}
\half(\mT_1\mQ\mT_1+\mE\mT_1\mQ\mT_1\mE)=
\Matr{1-\nu_1 & 0 & \mu_c^* & 0 \\
            0 & 1-\nu_1 & 0 & \mu_c \\
           \mu_c & 0 & 1-\nu_2 & 0 \\
            0 & \mu_c^* & 0 & 1-\nu_2}
\end{equation}
exceed unity.
Therefore, if $G>0$ so that the restrictions of (\ref{eq:2Dpos,tmQ})
are obeyed, $G\ptrans$ is positive if
\begin{equation}\label{eq:2Dsep.2}
\nu_1\nu_2\geq\mu_c^*\mu_c\phstar\;.
\end{equation}

Gaussians that are P-representable,
\begin{equation}
  G=\int\!(\D\mz)\; \sqrt{\det\mP}\,\Exp{-\half\mz\adj\mP\mz}
:\Exp{-(a_1\adj-z_1^*)(a_1\phadj-z_1\phstar)}
\Exp{-(a_2\adj-z_2^*)(a_2\phadj-z_2\phstar)}:
\nonumber\\
\end{equation}
(with $\mP>0$), are separable by construction.
\emph{Local} unitary transformations, $U_{\rm loc}=U_1U_2$ with
\begin{equation}
\mU_1=
\Matr{\Exp{\I\phi_1}\cosh\theta_1 &\Exp{\I\varphi_1}\sinh\theta_1 & 0 & 0 \\
    \Exp{-\I\varphi_1}\sinh\theta_1 & \Exp{-\I\phi_1}\cosh\theta_1 & 0 & 0 \\
            0 & 0 & 1 & 0  \\ 0 & 0 &\quad 0\quad &\quad 1 \quad}
\end{equation}
and
\begin{equation}
\mU_2=\Matr{\quad 1 \quad & \quad 0 \quad & 0 & 0 \\ 0 & 1 & 0 & 0\\
      0 & 0 & \Exp{\I\phi_2}\cosh\theta_2 &\Exp{\I\varphi_2}\sinh\theta_2 \\
      0 & 0 & \Exp{-\I\varphi_2}\sinh\theta_2 & \Exp{-\I\phi_2}\cosh\theta_2 }
\,,
\end{equation}
turn such a P-representable Gaussian into other separable Gaussians,
which may or may not be P-representable themselves.
All separable Gaussians can be constructed in this way,
and as a consequence the Peres criterion is not only necessary,
it is indeed sufficient.

Although this is reasonably obvious, the explicit construction of the unitary
transformation that does the job is technically more involved than the
analogous problem in one dimension that we treated in
Sec.~\ref{sec:1D-Prepr}.
We shall, therefore be content with the following general remarks,
and present an explicit example in Sec.~\ref{sec:antiEPR} below.

The separability of a Gaussian state can be investigated by studying
the existence of a local transformation $U_{\rm loc}=U_1U_2$ that
maps the  general Gaussian characteristic function
(\ref{eq:2DdefC}) into a P-representable Gaussian characteristic
function
\begin{equation}\label{eq:2DPrep}
U_{\rm loc}: \mC \rightarrow \mC_{\mathrm{P}}
\end{equation}
which has to satisfy the condition
\begin{equation}\label{eq:Prepcondition}
\mC_{\mathrm{P}} -\thalf \mI > 0\,.
\end{equation}
This says that the four eigenvalues of $\mC_{\mathrm{P}} -\thalf \mI$ are
strictly positive.
As shown in Refs.~\refcite{Duan+al00} and \refcite{Simon00},
this is indeed equivalent to the separability condition
(\ref{separability}).

\subsection{Examples}
We conclude this section on two-dimensional Gaussian operators 
by presenting a couple of examples, mostly generalizations of the historical
EPR Gaussian state.

\subsubsection{Pure Gaussian state}\label{sec:EPRpure}
We begin our examples with the simplest case of a pure Gaussian
state of a two-dimensional  harmonic oscillator. 
For pure states, we have $\Tr{G^2}=1$, and so (\ref{eq:trGG''}) implies 
that the Gaussian characteristic function must have a $\mC$ matrix 
(\ref{eq:2DdefC}) with
\begin{equation}\label{eq:det2DC}
  \det \mC = \frac{1}{16}\,.
\end{equation}
For convenience, we rewrite  (\ref{eq:2DdefC}) in the 
block form
\begin{equation}
 \mC=\matr{\mC_{11} & \mC_{12} \\ \mC_{21}&\mC_{22}}\,.
\label{eq:defpureC}
\end{equation}
For such pure states, the criterion for separability is simple, 
inasmuch as it is enough to study  the properties of
$\mC_{11}$ or $\mC_{22}$, which correspond to the reduced
one-dimensional parts of the two-dimensional harmonic oscillator.
If both $\mC_{11}$ or $\mC_{22}$ describe pure states, we say that
the joint state is not entangled. 
This amounts to
\begin{equation}
 \det\mC_{11}=\frac{1}{4} \ \ \mathrm{or} \ \ \det\mC_{22}=\frac{1}{4} \,.
\end{equation}
For Gaussian pure states this separability condition 
is much simpler than the general condition of Sec.~\ref{sec:2D-SepCrit} 
for mixed Gaussian states.

If we restrict the  parameters that characterize
a pure Gaussian state to real numbers only, and continue to ignore
linear shifts, such a state is described by a wave function of the form
\begin{equation}\label{eq:psi2}
 \Psi(\mq) = \left(\frac{\det \mD}{\pi^{2}}\right)^{\frac{1}{4}}\,
\Exp{-\half\mq\adj\mD\mq}
\end{equation}
in the position representation ($\mq=\row{q_{1},q_{2}}\adj$). 
Here we meet the $2\times2$ matrix 
\begin{equation}
 \mD =\mD\adj =\matr{\alpha & \gamma \\ \gamma & \beta} 
\quad \mbox{with} \quad \det\mD= \alpha\beta-\gamma^{2}\,,
  \label{eq:mDdef}
\end{equation}
and $\mD>0$, that is
\begin{equation}
  \alpha+\beta>\sqrt{(\alpha-\beta)^2+4\gamma^2}\,,
\end{equation}
is needed for the normalizability of the Gaussian wave function 
(\ref{eq:psi2}).

Then, the $\mC$ matrix of the resulting
Gaussian characteristic function (\ref{eq:defpureC}) has the blocks
\begin{eqnarray}
\label{eq:defC11}
\mC_{11}&=&\spmatr{\displaystyle\frac{\alpha}{4}+\frac{\beta}{4\det \mD}
&\displaystyle\frac{\alpha}{4}-\frac{\beta}{4\det \mD}\\[2ex]
\displaystyle\frac{\alpha}{4}-\frac{\beta}{4\det \mD}
&\displaystyle \frac{\alpha}{4}+\frac{\beta}{4\det \mD}
}\,,
\nonumber\\[2ex]
\mC_{22}&=&\spmatr{\displaystyle\frac{\beta}{4}+\frac{\alpha}{4\det \mD}
&\displaystyle\frac{\beta}{4}-\frac{\alpha}{4\det \mD}\\[2ex]
\displaystyle\frac{\beta}{4}-\frac{\alpha}{4\det \mD}& 
\displaystyle\frac{\beta}{4}+\frac{\alpha}{4\det \mD} }
\,,\nonumber\\[2ex]
 \mC_{12}=\mC_{21}&=&\frac{\gamma}{4}
\spmatr{\displaystyle1-\frac{1}{\det\mD} 
& \displaystyle1+\frac{1}{\det\mD}
  \\[2ex] 
\displaystyle1+\frac{1}{\det\mD} & \displaystyle1-\frac{1}{\det\mD}}\,.
\end{eqnarray}
It is easy to check that a $\mC$ matrix with these blocks
satisfies the purity condition (\ref{eq:det2DC}).

The separability condition $\det\mC_{11}=\frac{1}{4}$ amounts to
$\det \mD = \alpha\beta$, so that the state is
separable (not entangled) if and only if $\gamma=0$, 
as it should be expected for pure states. 
For $\gamma\neq 0$, all pure states represented by the Gaussian 
wave function (\ref{eq:psi2}) are entangled.

This wave function  defines a Gaussian operator that is a projector,
\begin{equation}
    G= \proj{\Psi}\,,
\end{equation}
where
\begin{equation}
\ket{\Psi} =(\det \mQ)^{\frac{1}{4}}
\,\Exp{-\half\mu_1{a_{1}\adj}^2} \Exp{
-\half\mu_2{a_{2}\adj}^2}\Exp{-\mu_c a_{1}\adj a_{2}\adj
}\,\ket{0,0}\,
\end{equation}
with
\begin{eqnarray}
\mu_1 &=& \displaystyle\frac{\det\mD+\alpha-\beta-1}{\det\mD+\alpha+\beta+1} 
\,,\nonumber\\[1ex]
\mu_2 &=&\displaystyle\frac{\det\mD-\alpha+\beta-1}{\det\mD+\alpha+\beta+1} 
\,,\nonumber\\[1ex]
\mu_c &=&\displaystyle \frac{2\gamma}{\det\mD+\alpha+\beta+1} 
\end{eqnarray}
being the non-zero parameters in (\ref{eq:2DmQ}), and
\begin{equation}
\det\mQ = \frac{16\det\mD}{(\det\mD+\alpha+\beta+1)^{2}}
\end{equation}
follows.
Therefore, such a Gaussian wave function
corresponds to   two squeezed  one-dimensional harmonic
oscillators, correlated with a strength that is characterized by
the parameter $\gamma$.

\begin{figure}
\centering
{\includegraphics[scale=1.35]{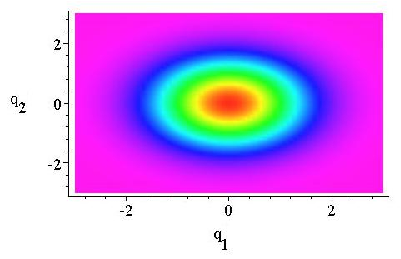}\hspace{0.8cm}
\includegraphics[scale=1.35]{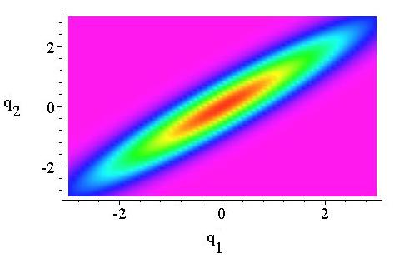}} 
\caption{\label{fig2}%
Contour plot of the smoothed EPR wave function (\ref{psiEPR})
for $\barr{n}=0$ (not entangled, on the left) 
and $\barr{n}=1$ (entangled, on the right).
}
\end{figure}

As a further simplification, and also as an important physical application
related to the EPR wave function (\ref{epr}), 
let us investigate a correlated state that is not squeezed. 
We obtain such a state by choosing the parameters in accordance with
\begin{equation}
\alpha=\beta \quad\mbox{and}\quad \det\mD =1\,,\enskip\mbox{so that}\quad 
\gamma^{2} =\alpha^{2} -1\,.
\end{equation}
Such a Gaussian state is characterized by a single real parameter,
for which $\barr{n}$ in
\begin{equation}
\label{eq:pureparameter}
    \alpha=\beta= 1+2\,\barr{n} \quad\mathrm{and} \quad 
|\gamma| = 2\sqrt{\barr{n}\,(\barr{n}+1)}
\end{equation}
is a convenient choice.
Upon opting for $\gamma=-|\gamma|$, the wave function (\ref{eq:psi2})
is
\begin{equation}
\label{psiEPR} \Psi(\mq) =\pi^{-\half}\,\exp\left(
-(\barr{n}+\thalf)(q^2_1+q^2_2) +2
    \sqrt{\barr{n}\,(\barr{n}+1)}\,q_1q_2\right)\,.
\end{equation}
For large $\barr{n}$ it reduces to
\begin{equation}
    \Psi(\mq)
    \sim \Exp{-\barr{n}\,(q_{1}-q_{2})^{2}}
\end{equation}
and becomes
\begin{equation}
   \Psi(\mq)  \sim \delta(q_{1}-q_{2})
\end{equation}
in the limit $\barr{n}\to\infty$.
We recognize here the famous historical wave function (\ref{epr}) of two
correlated particles introduced by the EPR trio. 
This wave function cannot be  normalized, and thus represents an
overidealized situation. 
By contrast, the wave function (\ref{psiEPR}) is a smoothed,
normalized version of (\ref{epr}) and refers to a real physical situation.
For illustration, we show this smoothed wave function in Fig.~\ref{fig2} 
for two different values of $\barr{n}$.

\subsubsection{Bell states}
For entangled spin-$\thalf$ states, a special role is played by
the entangled basis of the four Bell states. For the
two-dimensional harmonic oscillator it is possible to introduce a
continuous-variable generalization of Bell's states. Such a
complete set of states consists of particular Gaussian operators.

Following the example (\ref{eq:WParity}) of the singular Wigner
function for a one-dimensional harmonic oscillator, we write for
the two-dimensional case a singular Wigner function in the form
of a product of two two-dimensional Dirac delta functions,
\begin{equation}\label{eq:WEPR}
W(\mz)=\delta(\mz_{1}-\mz_{0})\delta(\mz_2+\mz_{0}^*)
\quad\mbox{with}\quad\mz=\col{\mz_1\\ \mz_2}\enskip\mbox{and}\enskip
\mz_0=\col{z_0\\z_0^*}\,,
\end{equation}
where the two-dimensional $\delta$-function $\delta(\mz)$ 
is as defined in (\ref{eq:delta-def}) 
and $z_{0}$ is an arbitrary complex number. 
The corresponding Gaussian operator,
\begin{eqnarray}\label{eq:GBell}
G(\mz_{0})
& =& \ : \, \Exp{-a_{1}\adj a_{1} -a_{2}\adj a_{2} +a_{1}
a_{2} + a_{1}\adj a_{2}\adj + z_{0}(a_{2}- a_{1}\adj)
+z_{0}^*(a_{2}\adj- a_{1})}\, : 
\nonumber\\&=&\proj{\mz_{0}}\,,
\end{eqnarray}
projects on the ``continuous Bell state'' specified by the ket
\begin{eqnarray}\label{eq:Bellstate}
\ket{\mz_{0}}
& =&  \Exp{ -\thalf|z_{0}|^{2} +z_{0}^*a_{2}\adj
- z_{0}a_{1}\adj +a_{1}\adj a_{2}\adj}\, \ket{0,0}\nonumber\\
&=&\Exp{\thalf|z_{0}|^{2}+(a_1\adj-z_0^*)(a_2\adj+z_0)}\,\ket{0,0}
\,,
\end{eqnarray}
which obtains from $\ket{\mz_0=[0,0]\trans}$ 
by a simultaneous unitary shift of $\ma_1$ by
$\mz_0$ and $\ma_2$ by $-\mz_0^*$, respectively.
The completeness relation
\begin{equation}
    \int\!(\D\mz_{0})\,\proj{\mz_{0}} = \mI
\end{equation}
is easily demonstrated with the aid of the Fourier--Gauss formula 
(\ref{eq:FourGauss}).

\subsubsection{Mixed EPR states}\label{sec:mixedEPR}
\begin{figure}
\centering{\includegraphics[angle=-90,scale=0.3]{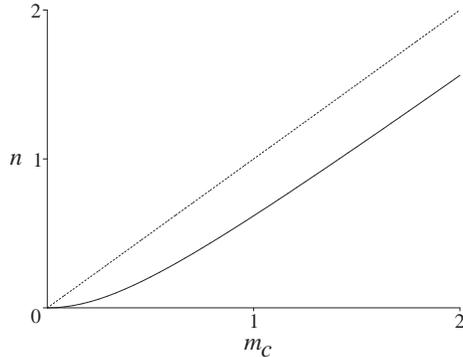}}
\caption{\label{fig3}%
Criteria for positivity and separability for the mixed EPR state specified by
(\ref{eq:Q1}). 
In this plot of $n$ versus $m_{c}$, the states associated with points on and
above the dashed line are separable, whereas points on and above the solid
line mark positive Gaussians, with pure states obtaining for $n,m_c$ values
on the solid line.}
\end{figure}

In this example the matric $\mC$ of the Gaussian characteristic
function is of the form
\begin{equation}
\label{eq:C1}
\mC = \Matr{n+\thalf & 0 & 0& m_{c}\\
0 & n+\thalf & m_{c}& 0\\0 & m_{c} & n+\thalf& 0\\ m_{c} & 0& 0
&n+\thalf }\,,
\end{equation}
so that  $n$ ($=n_1=n_2$) and $m_{c}$ are the only non-zero parameters in
(\ref{eq:2DdefC})
and, in addition, we take $m_c$ to be real. 
The corresponding $\mQ$ matrix of (\ref{eq:2DmQ}) is
\begin{equation}
\label{eq:Q1}
\mQ = \frac{1}{(n+1)^{2}-m_{c}^{2}}\Matr{n+1 & 0 & 0& m_{c}\\
0 & n+1 & m_{c}& 0\\0 & m_{c} & n+1& 0\\ m_{c} & 0& 0 &n+1 }\,,
\end{equation}
and the positivity criterion (\ref{eq:2Dpos,tmQ}) amounts to
\begin{equation}
\label{eq:po1}
 n(n+1)\geq m_{c}^{2}\,.
\end{equation}
This is less stringent than the condition for separability,
\begin{equation}
\label{eq:se1}
    n \geq |m_{c}|\,,
\end{equation}
that follows from (\ref{eq:2Dsep.2}).
These matters are illustrated in Fig.~\ref{fig3}.

Note that the Gaussian operator specified by (\ref{eq:C1}) represents a mixed
state in general. Only if
\begin{equation}
\det \mC = \left[(n+\thalf)^{2}-m_{c}^{2}\right]^{2} = \frac{1}{16}
\end{equation}
holds, the state is a pure state, which is the border case of the positivity
criterion (\ref{eq:po1}).
With the identification $\gamma= 2m_{c}$ and $n=\barr{n}$, we then recover 
the EPR pure state of Sec.~\ref{sec:EPRpure}, as we should.

We further note that the remarkably simple 
separability condition (\ref{eq:se1})
can be expressed in terms of Bohr's variables (\ref{bohr}) as
\begin{equation}
\label{bohrsepara}
 \expect{\mathcal{P}^2_1} + \expect{\mathcal{Q}^2_2} \geq 1 \quad
 \mathrm{and} \quad \expect{\mathcal{P}^2_2} +
 \expect{\mathcal{Q}^2_1} \geq 1\,.
\end{equation}
These relations have important physical implications when the
separability of the EPR state is put to an experimental test.
It is then enough to measure the variances of the Bohr
variables and check if the conditions (\ref{bohrsepara}) are
satisfied. 
One should remember, however, that 
(\ref{bohrsepara}) applies only to the very special state
characterized by the $\mC$ matrix of (\ref{eq:C1}). 
As we will see in the next examples, Gaussian
states different from this generalization of the EPR state 
will lead to a separability condition that will be different in form.

\subsubsection{Anti-EPR states}\label{sec:antiEPR}
We generalize the previous example by adding anti-EPR terms to the 
matrix $\mC$ of the Gaussian characteristic function in (\ref{eq:C1}),
so that now $\mC$ is of the form
\begin{equation}
\label{eq:C2}
\mC = \Matr{n+\thalf & 0 & m_{s}& m_{c}\\
0 & n+\thalf & m_{c}& m_{s}\\m_{s} & m_{c} & n+\thalf& 0\\ m_{c} &
m_{s}& 0 &n+\thalf }
\end{equation}
and is characterized by  the three real parameters $n$, $m_{c}$, and $m_{s}$. 
According to (\ref{eq:C->TCT}), the partially transposed Gaussian 
has $m_{c}$ and $m_{s}$ interchanged, 
and because of this property we call the
$m_{s}$ terms the anti-EPR contribution. 
The $\mQ$ matrix (\ref{eq:2DmQ}) of such a squeezed Gaussian operator 
has these non-zero elements:
\begin{eqnarray}
&\displaystyle
\nu_1=\nu_2=1-\frac{(n + 1)[(n+1)^{2} - m _{s}^{2} - m_{c}^{2}]}{d}\,,\quad
\mu_1=\mu_2=-\frac{m_{s}m_{c}(n+1)}{d}\,, &
\nonumber\\[1ex]
&\displaystyle
\mu _{s}=-m_{s}\frac{(n + 1)^2 -m_{s}^2+ m_{c}^2}{d}\,,\quad
\mu_{c}= m_{c}\frac{(n + 1)^2 +m_{s}^2- m_{c}^2}{d}  \,,&
\end{eqnarray}
where $d=[(n+1)^2-(m_{s}+m_{c})^2][(n+1)^2-(m_{c}-m_{s})^2]$.
As a consequence, the positivity condition (\ref{eq:2Dpos,tmQ}) reads here
\begin{equation}
\label{eq:po1a}
 n(n+1) - 2m_{s}(n+\thalf) +m_{s}^2 -m_{c}^{2} \geq 0\,,
\end{equation}
and the separability criterion (\ref{eq:2Dsep.2}) amounts to
\begin{equation}
\label{eq:se1a} n (n+1)- 2m_{c}(n+\thalf) +m_{c}^{2}- m_{s}^{2}\geq0\,.
\end{equation}
For $m_{s}=0$, we recognize the particular case of the mixed  EPR
state discussed in Sec.~\ref{sec:mixedEPR}. 

\begin{figure}
\centering{
\includegraphics[angle=-90,scale=0.29]{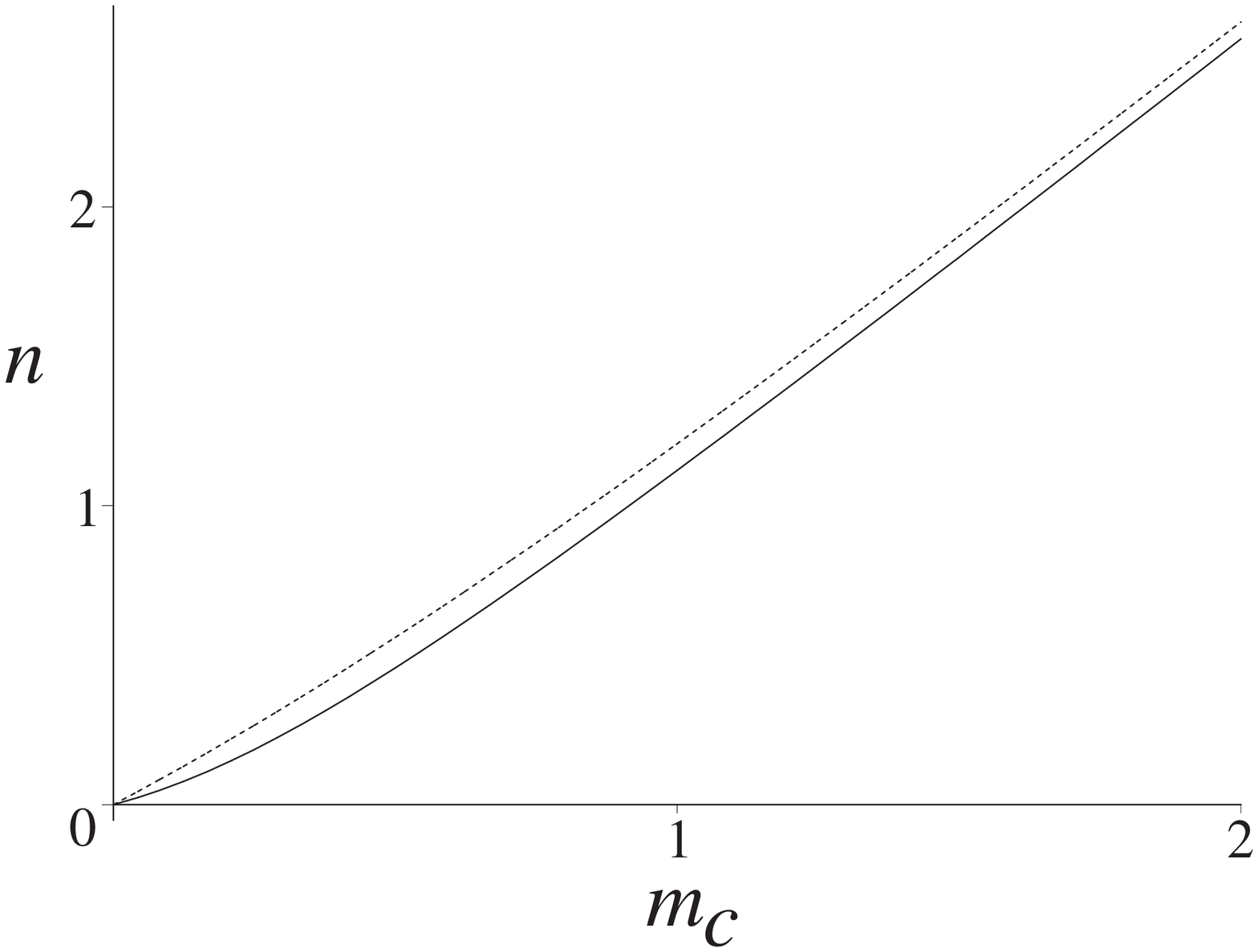}\hspace{0.8cm}
\includegraphics[angle=-90,scale=0.29]{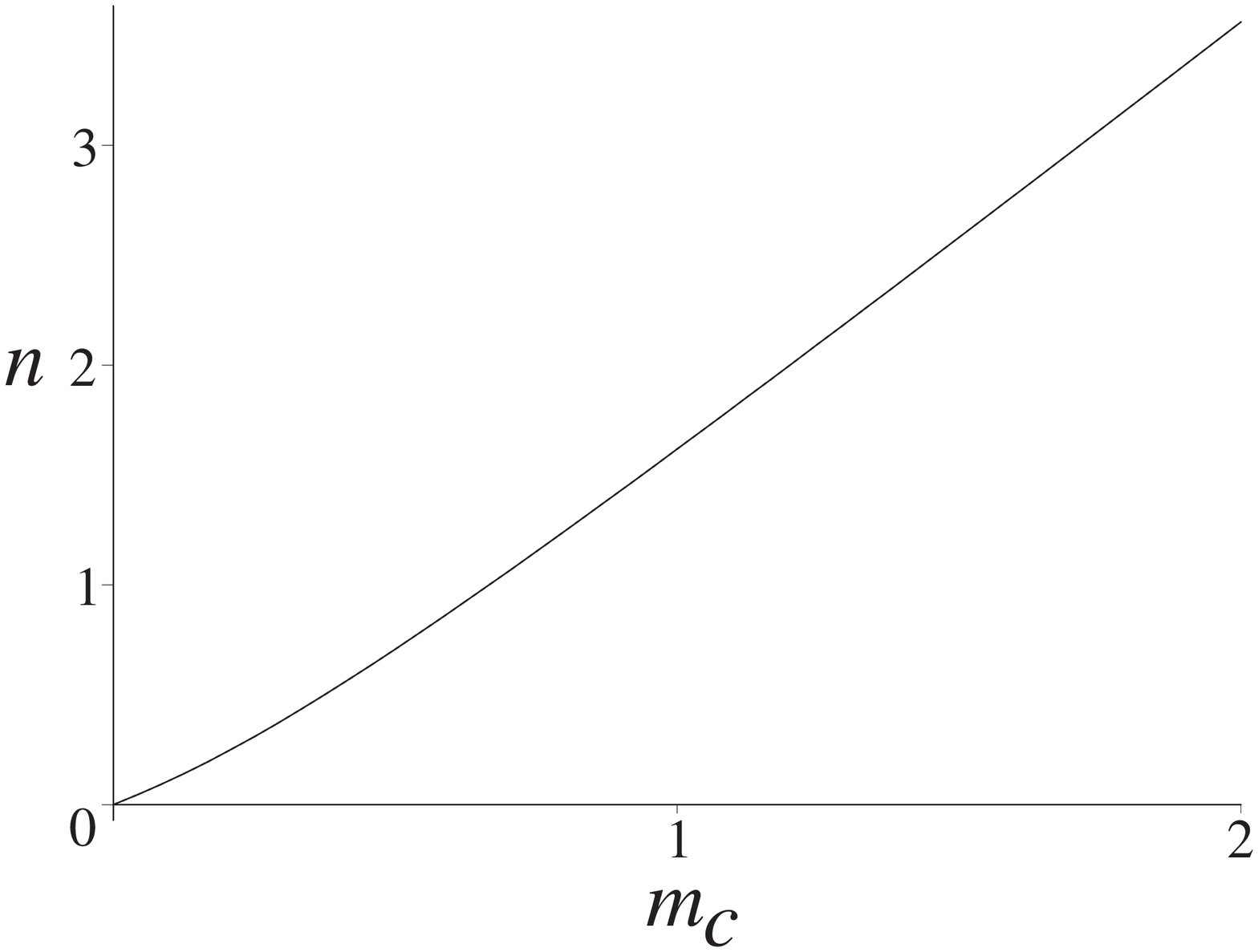}
}
\caption{\label{fig4}%
The positivity and separability criteria (\ref{eq:po1a}) 
and  (\ref{eq:se1a}) 
bound the region of non-separable anti-EPR Gaussian states (solid and dashed
curves, respectively). 
The left plot corresponds to $m_{s}=\half m_{c}$, 
the right plot to $m_s= m_{c}$.}
\end{figure}

In Fig.~\ref{fig4}, we depict the regions for the parameters $n$ and $m_c$ for
which the anti-EPR Gaussian states are non-separable, for two ratios of 
$m_s$ to $m_c$.
These regions are bounded
by the curves defined by the positivity condition (\ref{eq:po1a}) 
and the separability criterion (\ref{eq:se1a}). 
Note that the right plot corresponds to $m_s= m_{c}$, so that 
there the region of interest has no area at all, which means that 
such a state is never non-separable.

We conclude this example by the explicit construction of
the local transformation that maps the Gaussian state associated with
(\ref{eq:C2}) onto a Gaussian P-representable state, that is
\begin{equation}
U_{\rm loc}(\theta): \mC \rightarrow
\mC_{\mathrm{P}}(\theta)=\Matr{N+\thalf & M & M_{s}& M_{c}\\
M & M+\thalf & M_{c}& M_{s}\\M_{s} & M_{c} & N+\thalf& M\\ M_{c} &
M_{s}& M &N+\thalf }
\end{equation}
in terms of the $\mC$ matrix of (\ref{eq:C2}).
Because we have selected real parameters in $\mC$, 
the local transformation is labeled by a single parameter $\theta$. 
The transformed matrix $\mC_\mathrm{P}$ belongs to
a P-representable Gaussian operator if the
conditions
\begin{equation}
\label{eq:cond1}
 N\pm M \geq |M_{c}\pm M_{s}|
\end{equation}
are obeyed.
Here we have
\begin{eqnarray}
 N\pm M &=& (n+\thalf)\Exp{\mp 2\theta} - \thalf\,, \nonumber\\
|M_{c}\pm M_{s}|&= &\Exp{\mp2\theta}|m_{c}\pm m_{s}|\,.
\end{eqnarray}
The lower bounds in (\ref{eq:cond1}) are achieved if
\begin{eqnarray}
2M&=&|M_{c}+ M_{s}|-|M_{c}- M_{s}|  \,,\nonumber\\
2N&=&|M_{c}+ M_{s}|+|M_{c}- M_{s}|\,.
\end{eqnarray}
The first equality implies
\begin{equation}
(n+\thalf)( \Exp{- 2\theta}- \Exp{ 2\theta}) = |m_{c}+
m_{s}|\Exp{- 2\theta} - |m_{c}- m_{s}|\Exp{2\theta}
\end{equation}
or, more explicitly,
\begin{equation}
\label{eq:angle1}
 \Exp{4\theta} = \frac{n+\thalf - |m_{c}+
m_{s}|}{n+\thalf - |m_{c}- m_{s}|}
\end{equation}
for the transformation parameter $\theta$.
We see that for the two special cases of the EPR state ($m_{s}=0$)
and the anti-EPR case ($m_{c}=0$), we have $\theta=0$, and no
local transformation of the Gaussian characteristic function
(\ref{eq:C2}) is required. In these two cases the
P-representability can be studied directly by using (\ref{eq:C2}).

More generally, the conditions for P-representability 
that follow from (\ref{eq:cond1}) and (\ref{eq:angle1}) are
\begin{eqnarray}\label{eq:cond2}
 (n+ \thalf) -\thalf\Exp{-2\theta} &\geq &|m_{c}+  m_{s}| 
\,,\nonumber\\
 (n+ \thalf) -\thalf\Exp{2\theta} &\geq& |m_{c}- m_{s}|\,.
\end{eqnarray}
They are easily seen to be equivalent to the separability 
criterion (\ref{eq:se1a}).

\subsubsection{Squeezed EPR states}
As the last example of this tutorial we consider squeezed EPR states.
Their $\mC$ matrix (\ref{eq:2DdefC}) is specified by the three 
real and positive parameters $n,m_{c},m$ and is of the form
\begin{equation}
\mC = \Matr{n+\thalf & m & 0& m_{c}\\
m& n+\thalf & m_{c}& 0\\0 & m_{c} & n+\thalf& m\\ m_{c} & 0& m
&n+\thalf }\,.
\end{equation}
Note that this matrix differs from the one in (\ref{eq:C1}) 
by the additional squeezed terms associated with $m$ ($=m_1=m_2$). 
The $\mQ$ matrix (\ref{eq:2DmQ}) of such a squeezed Gaussian operator has the
non-zero elements
\begin{eqnarray}
&\displaystyle
\nu_1=\nu_2  = 1 - \frac{(n + 1)[(n+1)^{2} - m ^{2} - m_{c}^{2}]}{D}
\,,\quad
\mu_1=\mu_2
 = m\frac{(n+1)^{2}  - m^{2} + m_{c}^{2}}{D}\,,& \nonumber\\
&\displaystyle
\mu _{s} = \frac{2m m_{c}(n + 1)}{D}\,,\quad 
\mu _{c} =m_{c} \frac{(n+1)^{2}  - m_{c}^{2}  + m^{2}}{D} \nonumber \\
\end{eqnarray}
with $ D =[(n + 1)^2 - (m + m_{c})^2][(n + 1)^2 - (m - m_{c})^2]$.

Here the positivity condition (\ref{eq:2Dpos,tmQ}) requires
\begin{equation}
\label{eq:po}
 n(n+1) - (m_{c}+m)^{2} \geq 0
\end{equation}
and, according to the separability criterion(\ref{eq:2Dsep.2}), 
the squeezed EPR state is separable if
\begin{equation}
\label{eq:se} n (n+1)- 2m_{c}(n+\thalf) +m_{c}^{2}- m^{2}\geq 0
\end{equation}
holds.
For $m=0$, we recognize the particular case of the mixed  EPR
state discussed in Sec.~\ref{sec:mixedEPR}.

\begin{figure}
\centering{
\includegraphics[angle=-90,scale=0.29]{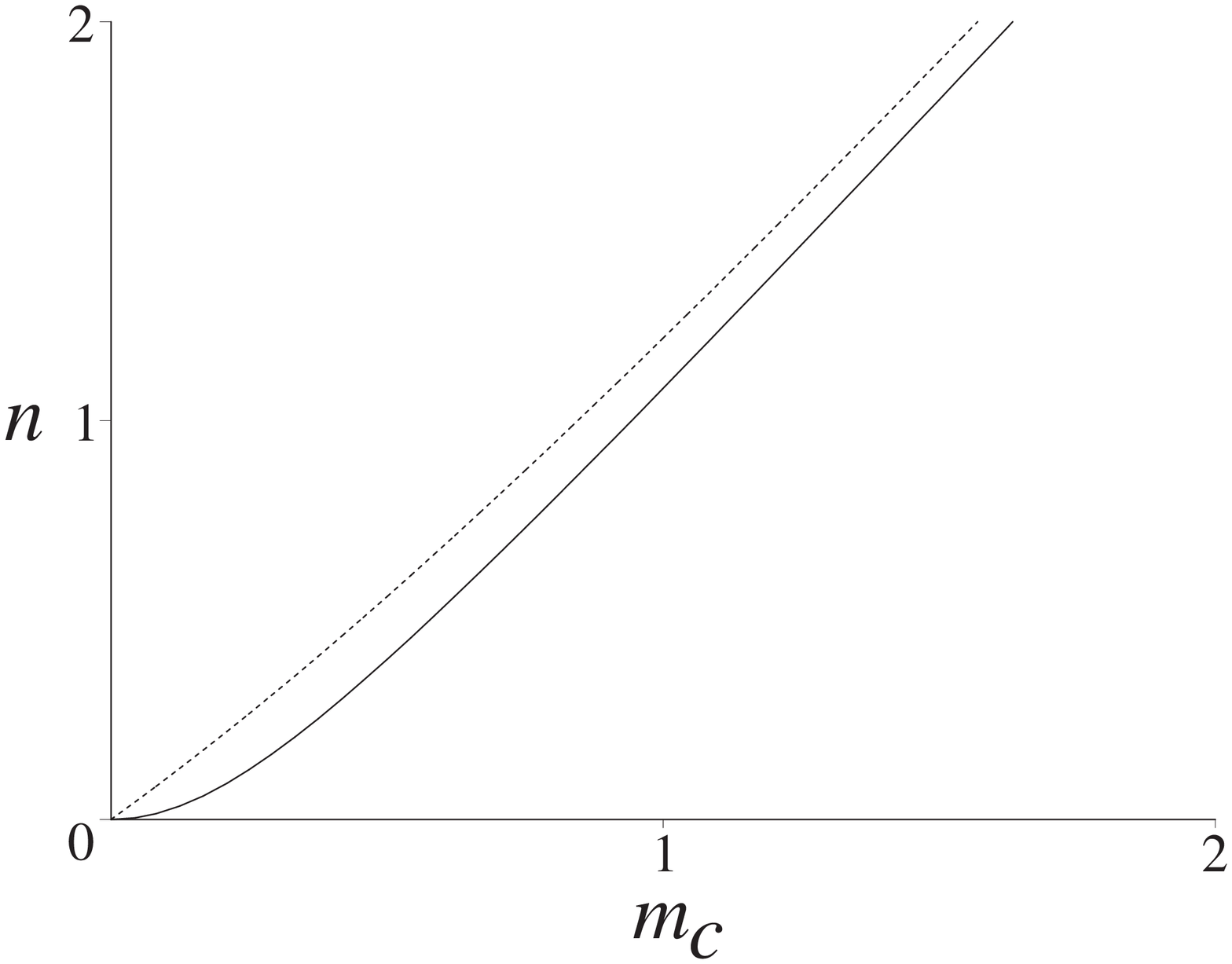}\hspace{0.8cm}
\includegraphics[angle=-90,scale=0.29]{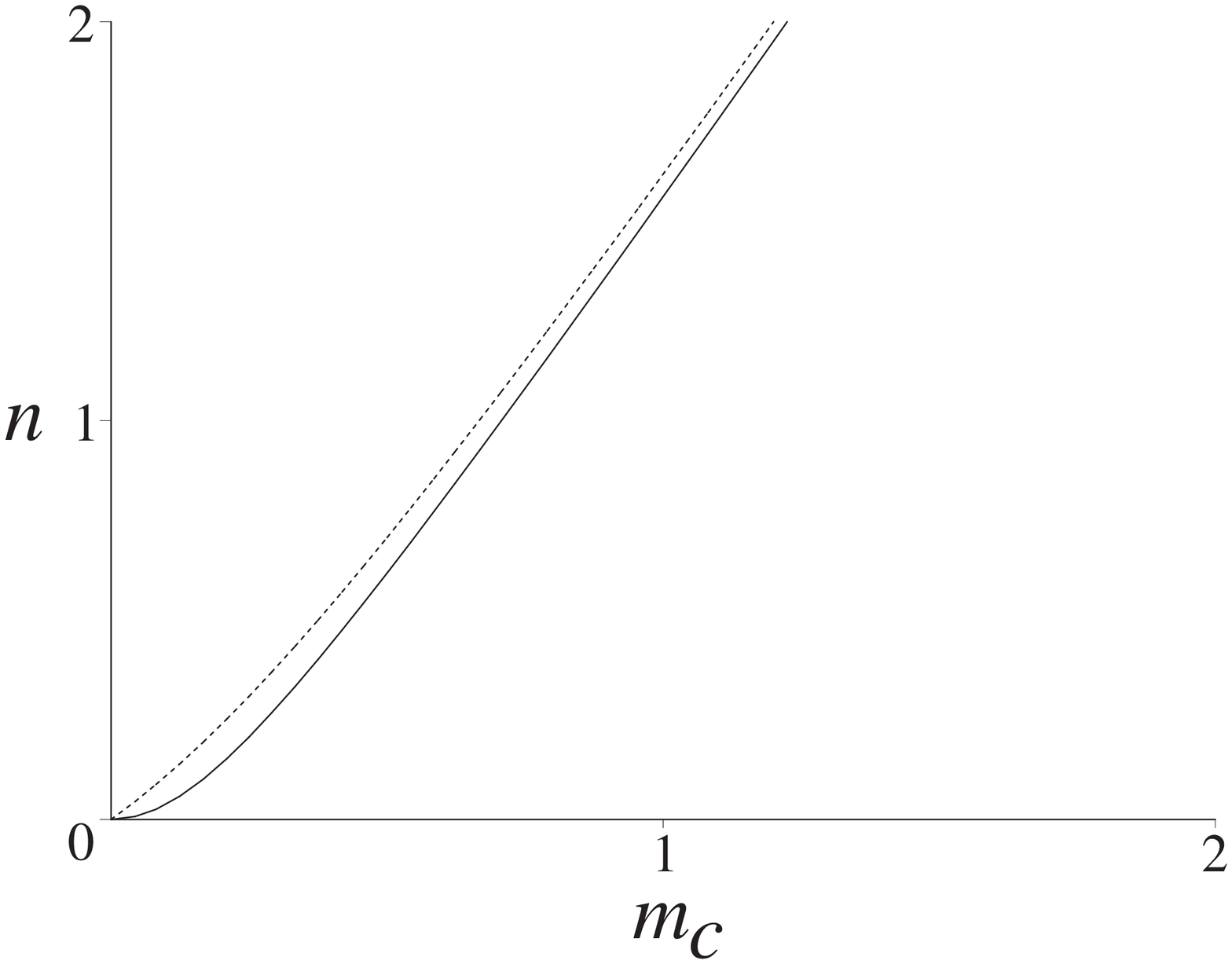}
}
\caption{\label{fig5}%
The positivity and separability criteria (\ref{eq:po}) 
and  (\ref{eq:se}) 
bound the region of non-separable squeezed 
EPR states (solid and dashed curves, respectively). 
The left plot corresponds to $m=\half m_{c}$, 
the right plot to $m= m_{c}$.}
\end{figure}

In Fig.~\ref{fig5}, we depict the regions for the parameters $n$ and $m_c$ for
which the anti-EPR Gaussian states are non-separable, for two ratios of 
$m$ to $m_c$.
These regions are bounded
by the curves defined by the positivity condition (\ref{eq:po}) 
and the separability criterion (\ref{eq:se}). 

\section{Concluding remarks}
The main goal of this tutorial is to provide, for a reader trained in Quantum
Optics or in Quantum Information Theory, a general and detailed description of
various mathematical tools used in the description of one-party and two-party
Gaussian states. 
For such states we presented a general parametrization of Gaussian states and
discussed the positivity criteria and the so-called P-representability of
positive Gaussian statistical operators. 
With the help of these techniques we showed that
it is possible to address the problem of quantum separability of two-party
Gaussian states using different-in-form, but mathematically equivalent, methods
based on the Heisenberg uncertainty relations, partial transposition, or the
P-representability.
The tutorial includes a number of simple, instructive examples that illustrate
the techniques and methods.

The one-dimensional Gaussian operators discussed in Sec. \ref{sec:1D}  
have been widely used in various application of squeezed states of light 
and  in the theory of
optical Gaussian beams  involving linear optical elements. 
For such one-dimensional
problems an important issue dealt with in the literature is the degree of
nonclassicality of one-mode Gaussian states of the quantum electromagnetic 
field. 
A general discussion of this problem has been given in Ref.~\refcite{Hillery}. 
In a different approach the nonclassicality  has been investigated with the
aid of the Uhlmann fidelity of two single-mode 
Gaussian states.\cite{Marian02}

Owing to their simplicity, 
Gaussian states have found many applications in Quantum
Information Theory. 
Examples of recent achievements that involve Gaussian states include
important problems such as entanglement key distribution,\cite{key} 
quantum error correction,\cite{error} 
entanglement distillation,\cite{distill} 
and quantum cloning.\cite{clone}
The separability and distillability of two-party Gaussian states are 
reviewed in Ref.~\refcite{giedke01}. 
The problem of quantum separability of three-party Gaussian states
and the entanglement criteria for all bipartite Gaussian states
have been addressed as well.\cite{three}
A number of issues related to the bound entangled 
Gaussian states,\cite{werner01}
and the optical generation of entangled Gaussian states\cite{generation} 
have also been addressed in the published literature.

Our tutorial is confined  to Gaussian states. 
The reader may wish to consult Ref.~\refcite{review00}, 
a ``primer'' on general composite quantum systems involving 
entangled states in discrete, finite-dimensional Hilbert spaces.

\section*{Acknowledgements}
We wish to thank Professor Herbert Walther for the hospitality and support
we enjoyed at the MPI f\"ur Quantenoptik where part of this work
was done. 
This work was partially supported by a KBN Grant 2 PO3B 021 23, 
and by the European Commission through the Research Training Network QUEST. 




\begin{thebibliography}{99}

\bibitem{Schroed27}
E. Schr\"odinger,  Ann.\ Phys.\ \textbf{79}, 361; 489; 734 (1926).

\bibitem{Heis27}
W. Heisenberg, \ZPhys \textbf{43}, 172 (1927).

\bibitem{Ga91} 
C. W. Gardiner, \textit{Quantum Noise} (Springer-Verlag, New York, 1991). 
See Chapter 4.45 for a discussion of the Gaussian operator.

\bibitem{EPR35}
A. Einstein, B. Podolsky, and N. Rosen, \PR \textbf{47}, 777 (1935).

\bibitem{Schroed35a}
E. Schr\"odinger,  \PCamPS \textbf{31}, 555 (1935).

\bibitem{Schroed35b}
E. Schr\"odinger, \JournalTitle{Naturwissenschaften} 
\textbf{23}, 807; 823; 844 (1935). 
English translation: J. D. Trimmer,
\JournalTitle{Proc.\ Am.\ Phil.\ Soc.}
\textbf{124}, 323 (1980); reprinted in \textit{Quantum Theory and Measurement},
J. A. Wheeler and W. H. Zurek, eds.\ 
(Princeton University Press, Princeton, 1983) pp.~152--167.

\bibitem{ekert}
A. Ekert, \textit{Quest for the True Origin of Entanglement}, 
http://cam.qubit.org/

\bibitem{bohr35} 
N. Bohr, \PR \textbf{48}, 698 (1935).

\bibitem{BaWo98} 
K. Banaszek and K. W\'odkiewicz, 
\PRA \textbf{58}, 4345 (1998).

\bibitem{Furusawa+al98}
A. Furusawa, J. L. S\o{}rensen, S. L. Braunstein, 
C. A. Fuchs, H. J. Kimble, and E. Polzik, 
\JournalTitle{Science} \textbf{282}, 706 (1998).

\bibitem{Ou+al92}
Z. Y. Ou, S. F. Pereira, H. J. Kimble, and K. C. Peng, 
\PRL \textbf{68}, 3663 (1992).

\bibitem{Braunstein+al01}
S. L. Braunstein, C. A. Fuchs, H. J. Kimble, and P. van Loock, 
\PRA \textbf{64}, 022321 (2001).

\bibitem{Werner89}
R. F. Werner, \PRA \textbf{40}, 4277 (1989).

\bibitem{Duan+al00}
L.-M. Duan, G. Giedke, J. I. Cirac, and P. Zoller, 
\PRL \textbf{84}, 2722 (2000).

\bibitem{Simon00}
R. Simon, \PRL \textbf{84}, 2726 (2000).

\bibitem{QO:tools}
W. Louisell, 
\textit{Quantum Statistical Properties of Radiation}
(Wiley, New York, 1973);  
H. J. Carmichael, 
\textit{Statistical Methods in Quantum Optics 1: 
Master Equations and Fokker-Planck Equations} 
(Springer-Verlag, Berlin, 1999); 
C. W. Gardiner and P. Zoller, 
\textit{Quantum Noise: A Handbook of Markovian and Non-Markovian 
Quantum Stochastic Methods With Applications to Quantum Optics}
(Springer-Verlag, Berlin, 2000).

\bibitem{QO:gen}
D. F. Walls and G. J. Milburn, 
\textit{Quantum Optics} (Springer-Verlag, New York, 1994); 
L. Mandel and E. Wolf, 
\textit{Optical Coherence and Quantum Optics}
(Cambridge University Press, Cambridge/UK, 2001).

\bibitem{QI:gen}
M. A. Nielsen and I. L. Chuang,
\textit{Quantum Computation and Quantum Information}
(Cambridge University Press, Cambridge/UK, 2000); 
G. Alber, Th.\ Beth, M. Horodecki, P. Horodecki,
R. Horodecki, M. R\"otteler, H. Weinfurter, R. Werner, and A. Zeilinger, eds.,
\textit{Quantum Information: An Introduction to Basic Theoretical Concepts 
and Experiments},
Springer Tracts in Modern Physics, Vol.~172 (Springer-Verlag, Berlin, 2001);
S. L. Braunstein and A. K. Pati, eds., 
\textit{Quantum Information Theory with Continuous Variables}
(Kluwer, Dordrecht, 2002).

\bibitem{Weyl27}
H. Weyl, \ZPhys \textbf{46}, 1 (1927).

\bibitem{Schwinger60}
J. Schwinger, \PNAS \textbf{46}, 570 (1960).

\bibitem{QM-SAM}
J. Schwinger, \textit{Quantum Mechanics. Symbolism of Atomic Measurements}
(Springer-Verlag, Berlin and Heidelberg, 2001).

\bibitem{Wigner32}
E. Wigner, \PR \textbf{40}, 749 (1932).

\bibitem{Moyal49}
J. E. Moyal, \PCamPS  \textbf{45}, 99 (1949).

\bibitem{Tatarskii83}
V. I. Tatarskii, \JournalTitle{Sov.\ Phys.\ Usp.} \textbf{26}, 311 (1983).

\bibitem{BalJen84}
N. Balazs and B. K. Jennings, \PhRep \textbf{104}, 347 (1984).

\bibitem{Hill+al84}
M. Hillery, R. F. O'Connell, M. O. Scully, and E. P. Wigner, 
\PhRep \textbf{106}, 121 (1984).

\bibitem{ScuZub97}
M. O. Scully and M. S. Zubairy, 
\textit{Quantum optics} (Cambridge University Press, Cambridge/UK, 1997).

\bibitem{Schleich01}
W. P. Schleich, \textit{Quantum optics in phase space} 
(Wiley-VCH, Weinheim, 2001).

\bibitem{Royer77}
A. Royer, \PRA \textbf{15}, 449 (1977).

\bibitem{Grossmann76}
A. Grossmann, \JournalTitle{Commun.\ Math.\ Phys.} \textbf{48}, 191 (1976).

\bibitem{bge89}
B.-G. Englert, \JPhysA  \textbf{22}, 625 (1989).

\bibitem{EngStWa93}
B.-G. Englert, N. Sterpi, and H. Walther, \OC \textbf{100}, 526 (1993).

\bibitem{BaWod96}
K. Banaszek and K. W\'odkiewicz, \PRL \textbf{76}, 4344 (1996); 
K. Banaszek, C. Radzewicz, K. W\'{o}dkiewicz, and J. S. Krasi\'{n}ski, 
\PRA \textbf{60}, 674 (1999).

\bibitem{LutDav97}
L. G. Lutterbach and L. Davidovich, 
\PRL \textbf{78}, 2547 (1997).

\bibitem{GarCalMosh80}
G. Garc\'\i{}a-Calder\'on and M. Moshinsky, 
\JPhysA \textbf{13}, L185 (1980).

\bibitem{EkKni90}
A. K. Ekert and P. L. Knight, 
\PRA \textbf{42}, 487 (1990).

\bibitem{EngFuPil02}
B.-G. Englert, S. A. Fulling, and M. D. Pilloff, 
\OC \textbf{208}, 139 (2002).

\bibitem{2Dentropy}
A. Serafini, F. Illuminati, and S. De Siena, 
eprint arXiv:quant-ph/0307073 (2003).

\bibitem{Peres96}
A. Peres, \PRL \textbf{77}, 1413 (1996).

\bibitem{we02}
B.-G. Englert and K. W\'odkiewicz, \PRA \textbf{65}, 054303 (2002).

\bibitem{Hillery}
M. Hillery, \PRA \textbf{39}, 2994 (1989).

\bibitem{Marian02}
P. Marian and T. A. Marian, \PRL \textbf{88}, 153601 (2002).

\bibitem{key}
T. C. Ralph, \PRA \textbf{61}, 010303R (2000); 
M. Hillery, \PRA \textbf{61}, 022309 (2000); 
M. D. Reid, \PRA \textbf{62}, 062308 (2000); 
S. F. Pereira, Z. Y. Ou, and H. J. Kimble, \PRA \textbf{62}, 042311 (2000); 
D. Gottesman and J. Preskill, \PRA \textbf{63}, 22309 (2001); 
N. J. Cerf, M. Levy, and G. van Assche, \PRA \textbf{63}, 052311 (2001); 
F. Grosshans and P. Grangier, \PRL \textbf{88}, 057902 (2002); 
Ch.\ Silberhorn, N. Korolkova, and G. Leuchs, \PRL \textbf{88}, 167902 (2002).

\bibitem{error}
S. L. Braunstein, \PRL \textbf{80}, 4084 (1998);
S .L. Braunstein, \JournalTitle{Nature (London)} \textbf{394}, 47 (1998).

\bibitem{distill}
G. Giedke, Duan Lu-Ming, J. L. Cirac, and P. Zoller, 
\JournalTitle{Quant.\ Inf.\  Comp.} \textbf{1}, 79 (2001); 
Wang Xiang-Bin, M. Keiji, T. Akihisa, \PRL \textbf{87}, 137903 (2001); 
G. Giedke, J. I. Cirac, \PRA \textbf{60}, 32316 (2002); 
J. Eisert, S. Scheel, and M. B. Plenio, \PRL \textbf{89}, 137903 (2002); 
J. Fiurasek, \PRL \textbf{89}, 137904 (2002); 
D. Bru\ss, J. I.  Cirac, P.  Horodecki, F. Hulpke, B. Kraus, 
M. Lewenstein, and A. Sanpera,  \JMO \textbf{49}, 1399 (2002).

\bibitem{clone}
N. J. Cerf, A. Ipe, and X. Rottenberg, \PRL \textbf{85}, 1754 (2000);
N. Cerf and S. Iblisdir, \PRA \textbf{62}, 040301 (2000); 
S. L. Braunstein, N. J. Cerf, S. Iblisdir, P. van Loock, and S. Massar, 
\PRL \textbf{86}, 4938 (2001).

\bibitem{giedke01}
G. Giedke, B. Kraus, Duan  Lu-Ming, P. Zoller, J. I. Cirac, and M. Lewenstein,
\JournalTitle{Fortschr.\  Phys.} \textbf{49}, 973 (2001).

\bibitem{three}
G. Giedke, B. Kraus, M. Lewenstein, and J. I. Cirac, 
\PRA \textbf{64}, 052303 (2001); 
G. Giedke, B. Kraus, M. Lewenstein, and J. I. Cirac, 
\PRL \textbf{87}, 167904 (2001).

\bibitem{werner01} 
R. F. Werner and M. M. Wolf, 
\PRL \textbf{86}, 3658 (2001).

\bibitem{generation} 
P. Marian, T. A. Marian, H. Scutaru, 
\JPhysA \textbf{34}, 6969 (2001); 
M. S. Kim, J. Lee, W. J. Munro,  \PRA \textbf{66}, 30301 (2002).

\bibitem{review00}
M. Lewenstein, D. Bru\ss, J. I.  Cirac, B. Kraus, M. Ku\'s, J. Samsonowicz, 
A. Sanpera, and R. Tarrach, 
\JMO \textbf{47}, 2481 (2000).

\end{thebibliography}
\end{document}